\documentclass[twocolumn]{aastex631}
\usepackage{booktabs}
\usepackage{filecontents,catchfile}
\usepackage{longtable}

\usepackage{CJKutf8}
\usepackage{amsmath}
\usepackage{cleveref}

\newcommand{\um}{\,\micron}

\newcommand{\molH}{H\ensuremath{_\mathrm{2}}}

\newcommand{\PAHFIT}{\texttt{PAHFIT}}

\newcommand{\LIR}{\,L\ensuremath{_\mathrm{IR}}}
\newcommand{\Lsun}{\,L\ensuremath{_\mathrm{\odot}}}

\newcommand{\tauSi}{\ensuremath{\tau_\mathrm{Si}}}
\newcommand{\tauWater}{\ensuremath{\tau_\mathrm{H_{2}O}}}

\newcommand{\Robs}{\ensuremath{R_{\mathrm{obs},i}}}
\newcommand{\Rint}{\ensuremath{R_{\mathrm{int},i}}}

\newcommand{\paperII}{Paper~\textsc{ii}}

\renewcommand{\thefootnote}{\fnsymbol{footnote}}

\graphicspath{{./}{figures/}}

\received{\today}

\submitjournal{ApJ}

\shorttitle{Constraining the MIR attenuation curve using PAH}
\shortauthors{Lai et al.}

\begin{document}
\begin{CJK*}{UTF8}{bsmi}
\title{Spectroscopic Constraints on the Mid-Infrared Attenuation Curve: I - Attenuation Model using PAH Emission}

\correspondingauthor{Thomas S.-Y. Lai}
\email{ThomasLai.astro@gmail.com}

\author[0000-0001-8490-6632]{Thomas S.-Y. Lai (賴劭愉)}
\affil{IPAC, California Institute of Technology, 1200 E. California Blvd., Pasadena, CA 91125}
\affil{Ritter Astrophysical Research Center, University of Toledo, Toledo, OH 43606, USA}

\author[0000-0003-1545-5078]{J.D.T. Smith}
\affiliation{Ritter Astrophysical Research Center, University of Toledo, Toledo, OH 43606, USA}

\author[0000-0002-2541-1602]{Els Peeters}
\affiliation{Department of Physics \& Astronomy, The University of Western Ontario, London ON N6A 3K7, Canada}
\affiliation{Institute for Earth and Space Exploration, The University of Western Ontario, London ON N6A 3K7, Canada}
\affiliation{Carl Sagan Center, SETI Institute, 339 Bernardo Avenue, Suite 200, Mountain View, CA 94043, USA}

\author[0000-0002-8712-369X]{Henrik W.W. Spoon}
\affiliation{Cornell University, CRSR, Space Sciences Building, Ithaca, NY 14853, USA}

\author[0000-0002-9850-6290]{Shunsuke Baba (馬場俊介)}
\affiliation{Graduate School of Science and Engineering, Kagoshima University, 1-21-35 Korimoto, Kagoshima, Kagoshima 890-0065, Japan}

\author[0000-0001-6186-8792]{Masatoshi Imanishi (今西昌俊)}
\affiliation{National Astronomical Observatory of Japan, National Institutes of Natural Sciences (NINS), 2-21-1 Osawa, Mitaka, Tokyo 181-8588, Japan}

\author[0000-0002-6660-9375]{Takao Nakagawa (中川貴雄)}
\affiliation{Institute of Space and Astronautical Science (ISAS), Japan Aerospace Exploration Agency (JAXA), 3-1-1 Yoshinodai, Chuo-ku, Sagamihara, Kanagawa 252-5210,
Japan}



\begin{abstract}
We introduce a novel model to spectroscopically constrain the mid-infrared (MIR) extinction/attenuation curve from 3--17\um, using Polycyclic Aromatic Hydrocarbon (PAH) emission drawn from an AKARI-Spitzer extragalactic cross-archival dataset. Currently proposed MIR extinction curves vary significantly in their slopes toward the near-infrared, and the variation of the strengths and shapes of the 9.7\um\ and 18\um\ silicate absorption features make MIR spectral modeling and interpretation challenging, particularly for heavily obscured galaxies. By adopting the basic premise that PAH bands have relatively consistent intrinsic ratios within dusty starbursting galaxies, we can, for the first time, empirically determine the overall shape of the MIR attenuation curve by measuring the differential attenuation at specific PAH wavelengths. Our attenuation model shows PAH emission in most (U)LIRGs is unambiguously subjected to attenuation, and we find strong evidence that PAH bands undergo differential attenuation as obscuration increases. Compared to pre-existing results, the MIR attenuation curve derived from the model favors relatively gray continuum absorption from 3---8\um\ and silicate features with intermediate strength at 9.7\um\ but with stronger than typical 18\um\ opacity.

\end{abstract}



\section{Introduction} \label{sec:intro}
The selective attenuation of light by interstellar and intergalactic dust has a profound effect across a wide range of fields in astrophysics \citep[see][for reviews]{Draine2003, Salim2020}.  Considerable effort has been invested in understanding the wavelength dependence, feature set, parameterization, and variability of the extinction or attenuation curve; the latter being a more general analog of extinction in the context of multiple emission sources at varying depths. While in general, the attenuation curves in galaxies are relatively smooth, rising towards the ultraviolet, certain local features have a major impact on the interpretation of spectral features in their vicinity.  The most prominent and well-known attenuation feature is the 2175\AA\ bump, likely arising from resonant absorption by carbon grains \citep{Stecher1965, Mathis1994, Steglich2010}.  But at longer wavelengths, additional strong features appear, including the 10 and 18\um\ (O-)Si-O resonances in silicate-rich grains \citep{Woolf1969, vanBreemen2011}.  These features are often observed in starburst galaxies \citep[e.g.,][]{Rigopoulou1999} and (U)LIRGs \citep[e.g.,][]{Genzel1998, Tran2001, Spoon2004, Armus2004}. While the absolute level of attenuation is much lower at infrared wavelengths than in the UV, the 10\um\ silicate feature is actually \emph{more pronounced} than the 2175\AA\ UV absorption feature, with $A_\lambda$ rising above the underlying smooth continuum extinction by a factor of 2.4 \citep[compared to 1.4$\times$ for the 2175\AA\ feature; see][]{Hensley2021}. This salient contrast to the continuum has a major impact on the recovery of Polycyclic Aromatic Hydrocarbon (PAH) emission, with two of the strongest PAH complexes (at 7.7 and 11.3\um) flanking the 9.8\um\ resonant absorption band, and the longest PAH features at 17\um\ near the deepest part of 18\um\ silicate absorption. 

In addition, shortward of the silicate bands, the shape of the smooth attenuation continuum between 3---8\um\ departs substantially from the overall power-law decline, potentially flattening into a nearly gray curve \citep[e.g.,][]{Lutz1999, Indebetouw2005, Nishiyama2009}. Despite all of this complexity, much less is known about the morphology of the infrared attenuation curve from 3---25\um\ than at shorter wavelengths. In part, this is due to the reduced reddening at wavelengths longer than 5\um\ (often assumed, incorrectly, to be zero), and the lack of measurements, in particular, in regions outside the Milky Way, much less in diverse samples of galaxies. 

Measuring the shape of the extinction curve requires knowledge of the unreddened shape of the background radiation field via a fiducial model.  These can include pairs of stars of similar known spectral type \citep[e.g.,][]{Stecher1965, Massa1983, Gordon2021, Decleir2022}, ratios of lines with fixed known unreddened values (e.g. the H$\alpha$/H$\beta$ Balmer or other recombination line decrements; \citet{Moustakas2006}), and models of the UV continuum emission shape arising from a range of stellar populations \citep{Calzetti1994}. Based on JWST spectroscopy, the latter approach has recently been applied to high redshift galaxies to constrain the attenuation law at z$\sim$7---8 \citep{Markov2023, Witstok2023}, suggesting the presence of dust that gives rise to the 2175\AA\ bump in the first billion years of cosmic time. 

Here we propose a novel approach to assess the underlying shape of the mid-infrared (MIR) attenuation curve in galaxies --- using the approximate constancy of the vibrational PAH emission features in pure star-forming systems at near-solar metal abundance.  Since these PAH features are well-distributed throughout the range of wavelengths exhibiting the maximum variation of the MIR attenuation curve, such a method is able to place direct constraints on the attenuation at 3---17\um. In addition to these direct attenuation constraints that can be inferred from PAHs, in our companion paper (Lai et al. \textit{in prep.}; \paperII), we will present a detailed study on how \molH\ rotational lines can be used to further constrain the silicate absorption strength at 9.7\um. 



We organize this paper as follows: in \S2, differences between extinction and attenuation is explained, as well as a recap of the literature extinction curves used in this study. \S3 describes the sample, followed by an introduction to the PAH attenuation model and its derived results in \S4. We give caveats of the model in \S5 and discuss our results in \S6. Finally, we summarize our study in \S7. 


\section{Background}
Understanding the shapes and variations of dust attenuation is fundamental for the correct modeling of diverse physical properties of galaxies. Numerous studies have made clear that the adopted attenuation curve can strongly alter the derived physical properties of observed galaxies, e.g., star formation rate and stellar mass \citep{Kriek2013, Reddy2015, Shivaei2015, Salim2016}. In addition, interstellar dust modeling uses the extinction curve to constrain basic dust properties like sizes and compositions of grain particles \citep[e.g.,][]{Mathis1977, Desert1990, Weingartner2001, Jones2013, Hensley2021}. For mid-infrared spectral decomposition tools, such as \PAHFIT\ \citep{Smith2007}, unknown variations in the shape of the attenuation law directly impact the measurement of PAH band strengths because it complicates the process of defining the underlying continuum. \PAHFIT\ has demonstrated a successful approach to constrain the continuum in galaxies with low to moderate silicate optical depth (\tauSi) by scaling an adopted attenuation curve against the composite of multiple modified black bodies. For more obscured, dusty galaxies, however, this simple treatment is no longer viable. The spectra of some of these galaxies show a sharp kink and steepening of the continuum slope at $\sim$18\um, a feature for which the \PAHFIT\ adopted attenuation curve cannot adequately account. Most importantly, even if there is no universal attenuation law that applies in all environments, attempting to find a more realistic MIR attenuation curve that can be adopted in more galaxies is an important goal for correctly interpreting the MIR diagnostics derived from the spectral decomposition.

\subsection{Extinction vs. Attenuation}
The terms extinction and attenuation are essentially different, though they are sometimes used interchangeably in the literature. Extinction represents the photons being absorbed or scattered out of the line of sight by grains, so it is directly proportional to the column density along the way. In contrast, attenuation is more complex in the sense that it takes into account multiple sources with diverse star-dust geometry and considers light scattering back into the line of sight, and the composite effects of emitters at different depths \citep{Disney1989, Witt1996, Witt2000}. 


Extinction/attenuation curves at UV/optical wavelengths are often characterized by two main features: the UV-optical slope and the strength of the UV bump at 2175\,\AA\, \citep{Stecher1965}, which can vary drastically from galaxy to galaxy in simulations \citep{Narayanan2018}. The average attenuation curve in galaxies was first derived based on a number of UV-selected starburst galaxies, and the established correlation between the UV slope ($\beta$) and the nebular emission lines H$\alpha$ and H$\beta$ made the Balmer decrement an ideal method to probe the continuum attenuation by accessible observables \citep{Calzetti1994}. The Balmer decrement hinges on the fiducial, theoretical value expected for the unreddened H$\alpha$/H$\beta$ ratio \citep[2.88 in Case B recombination theory;][]{Hummer1987} and attributes deviations of observed ratios from this value to attenuation. 


The Milky Way extinction curve from the UV/optical has long been known to vary along different lines of sight, with changes that can largely be characterized by a single total-to-selective ratio, R$_\mathrm{V}$ = A$_\mathrm{V}$/E(B-V) \citep{Cardelli1989, Fitzpatrick2019}. Likewise, variations were also found in the IR extinction law along different lines of sight, suggesting that the extinction law may not be universal at IR wavelengths \citep{Fitzpatrick2009, Gao2009, Zasowski2009}. Early observations suggested that the MIR extinction curve at 3---8\um\ is a simple extension of the NIR power law, exhibiting a sharp minimum at $\sim$8\um\ before raising to the absorption peak of the 9.7\um\ silicate \citep[e.g.][]{Rieke1985, Bertoldi1999, Rosenthal2000}.  However, recent evidence suggests a flatter curve at wavelengths shortward of 8\um, first identified via hydrogen recombination lines toward the Galactic Center \citep{Lutz1996, Lutz1999}. Since then, numerous results have favored a flatter slope of the extinction curve \citep[e.g.,][]{Indebetouw2005, Nishiyama2006, Nishiyama2009, Jiang2006, Gao2009, McClure2009, Zasowski2009, Fritz2011, Wang2013, Xue2016, Gordon2021, Decleir2022, Gordon2023, Li2023}. The variation of the MIR slope can mainly be bracketed by the extinction curves calculated from dust models \citep[e.g.,][]{Weingartner2001} with R$_\mathrm{V}$=5.5 and R$_\mathrm{V}$=3.1, with a higher value of R$_\mathrm{V}$ representing a shallower slope, while a lower value of R$_\mathrm{V}$ a sharper minimum.

\begin{figure*}
  \centering
  \includegraphics[width=1\textwidth]{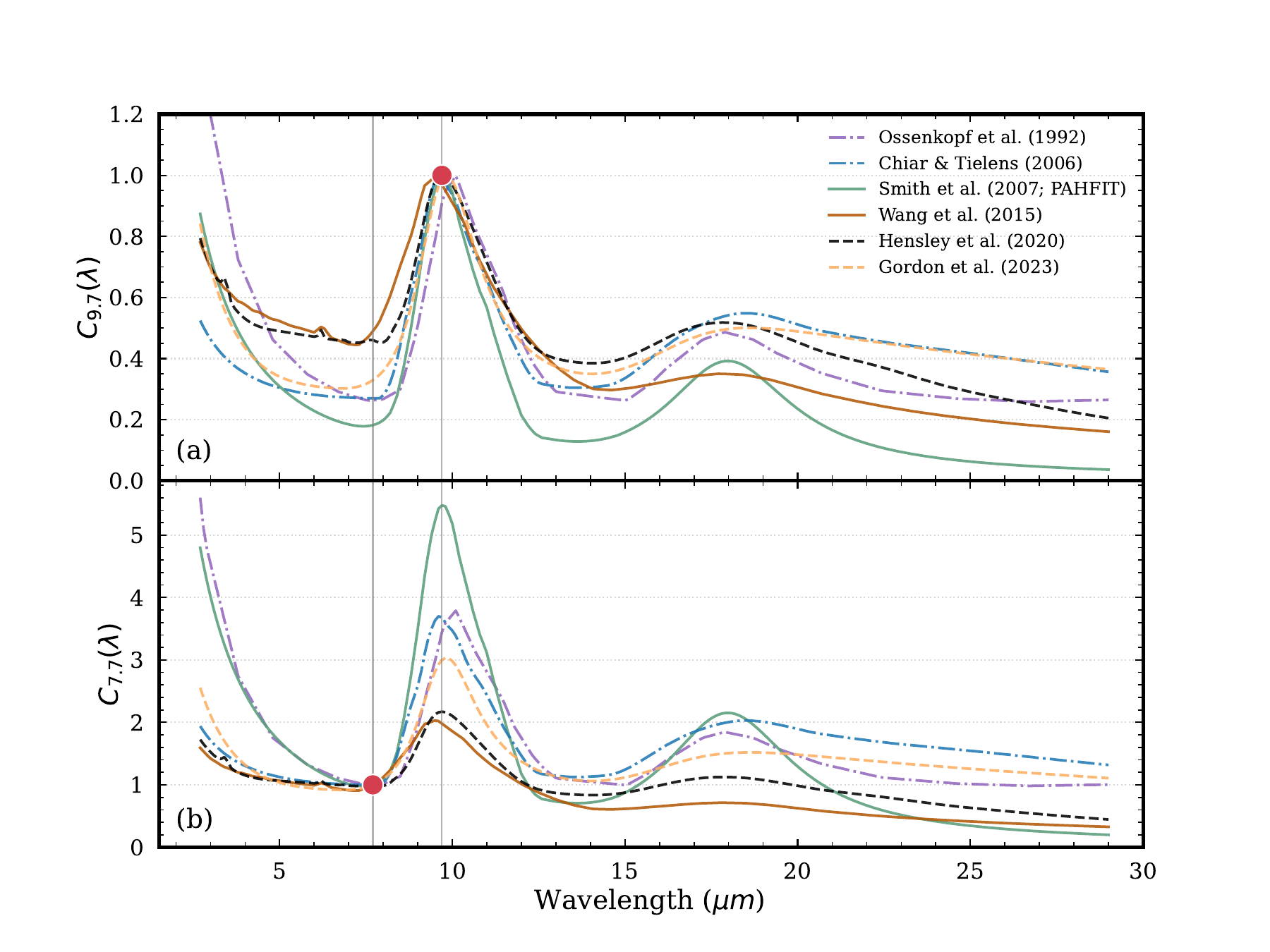}
  \caption{Comparison of the literature extinction curves in range of 3---29\um, including the extinction curve used in \PAHFIT\ \citep[][green]{Smith2007}, which is constructed on the one presented in \citet{Kemper2004}, along with other curves taken from \citet{Ossenkopf1992}, \citet{Chiar2006}, \citet{Wang2015}, \citet{Hensley2020}, and \citet{Gordon2023}. (a) The wavelength-dependent extinction curves are presented using the notation of $C_{9.7}({\lambda})$, with the subscript representing the wavelength used for the normalization of the curve. (b) Re-normalizing the curves to 7.7\um, the wavelength of the fiducial 7.7\um\ PAH band in the PAH attenuation model. This re-normalization is needed as the PAH attenuation model provides constraints on the differential attenuation between various PAH wavelengths (3.3, 6.2, 8.6, 11.3, 12.6, and 17\um) with respect to that at 7.7\um.
}
  \label{fig:extinction_curve_comparison}
\end{figure*}

The term ``extinction" captures the combined effects of scattering and absorption. In the near and mid-IR, where the average photon wavelength is large compared to typical grain sizes, scattering becomes negligible and absorption dominates the extinction \citep{Corrales2016}.  We focus here on constraining the \emph{attenuation} curve in the MIR based on AKARI+Spitzer extragalactic observations and put it in the same context with several proposed \emph{extinction} curves from the literature for comparison and explicitly consider different attenuation geometries to gauge the sensitivity of the results.

\subsection{MIR Extinction curve comparison}
\label{sec:extinction_curve_comparison}
The default extinction profile used in \PAHFIT, which ranges from 5---38\um, is a synthesized curve based on the 10\um\ silicate profile derived from \citet{Kemper2004} plus a broad Lorentzian absorption profile characterizing the 18\um\ feature that has a fixed ratio of peak absorption to the 9.7\um\ band, together with a raising power-law towards shorter NIR wavelengths \citep[][see Figure~\ref{fig:extinction_curve_comparison}(a)]{Smith2007}. Hereafter, S07 will be used to represent this PAHFIT-adopted extinction curve. The other extinction curve adopted in \PAHFIT\ is the profile constructed by \citet[][hereafter CT06]{Chiar2006}, derived from heavily obscured galactic WC-type Wolf-Rayet stars. This profile exhibits a flatter slope at 3--8\um\ compared to S07. Another profile with a slope at 3--8\um\ that is similar to S07 is the extinction curve of the cool, oxygen-rich silicates in \citet[][hereafter O92]{Ossenkopf1992}, which has been shown to match well with the deep silicate absorption found in ULIRGs \citep{Sirocky2008}. Since the development of \PAHFIT, more extinction curves with slightly different shapes have been proposed. For instance, \citet[][hereafter W15]{Wang2015} concluded that invoking micron-sized graphite grains in addition to the nano- and submicron-sized silicate-graphite-PAH model \citep{Draine2007} fits well with the observations that support a flat MIR extinction at 3--8\um\ \citep[e.g.,][]{Lutz1999}. \citet[][hereafter HD20]{Hensley2020} published a complete MIR extinction curve measured from the Galactic blue hypergiant Cyg OB2-12, showing a similar result to CT06 and W15 in terms of the shape of the extinction curve at 3--8\um. More recently, the MIR extinction curve has been modeled in an average over more than 10 diffuse sightlines with a functional form using the combination of two Lorentzian profiles by \citet[][hereafter G23]{Gordon2023}. This parameterized extinction curve appears to have a somewhat steeper slope than CT06, W15, and HD20, leaning more toward O92 and S07. 

To demonstrate the variations among these six measured IR extinction laws, we compare the \PAHFIT\ adopted extinction curves (S07) to profiles from O92, CT06, W15, HD20, and G23. In Figure \ref{fig:extinction_curve_comparison}(a), following the approach of \citet{Smith2007}, we normalize the extinction curves at 9.7\um\ and further introduce a term $C_{9.7}(\lambda)$ to represent the relative extinction at different wavelengths ($\lambda$), with respect to that at 9.7\um, as indicated by the subscript. In Fig. \ref{fig:extinction_curve_comparison}(b), all the curves are normalized at 7.7\um; the need for this re-normalization will be detailed in Section \ref{sec: Grid_model}. To illustrate, according to the notation, $C_{7.7}(9.7)$ is $\sim5.4$ in the case of S07. 

Three conspicuous differences can be seen among these curves: (i) the steepness of the curve between 3--8\um, (ii) the strength of the 9.7\um\ silicate feature, and (iii) the amount and shape of the extinction at 18\um. Most profiles, except CT06, exhibit strong extinction at $\sim$3\um\ when normalized at 9.7\um\ (in the $C_{9.7}(\lambda)$ space), but there is notable variation in the 3--8\um\ range, with some rising steeply towards shorter wavelengths, while others exhibiting ``gray" or flat extinction over this range. The difference between the strength of the 9.7\um\ silicate feature is more visible when normalized at 7.7\um, where S07 is the sharpest, followed by O92, CT06, and G23, with HD20 and W15 showing lesser contrast between 7.7 and 9.7\um. Knowing the shape of the attenuation curve in this region is crucial since it directly impacts measurements of the two flanking PAH bands at 7.7\um\ and the 11.3\um\ complex. The inconsistencies in the width, position, and strength of the 18\um\ silicate feature can also lead to challenges for recovering the 17\um\ PAH complex. The actual shape of the silicate feature plays an important role in the overall MIR spectral fitting of highly obscured sources with deep silicate absorption because the continuum assignment largely depends on whether the shape at the 10\um\ ``trough" matches well between the data and the employed attenuation profile.

\section{Sample and Observations}
\label{sec:sample}
The sample used in this study is taken from the PAH-bright galaxies in the AKARI-Spitzer Extragalactic Spectral Survey \citep[ASESS;][]{Lai2020}, which is a subset of the Infrared Database of Extragalactic Observables from Spitzer (IDEOS) catalog \citep{Hernan-Caballero2016, Spoon2022}. The PAH-bright sample primarily comprises (U)LIRGs, with \LIR$\geq10^{11}$\Lsun, accounting for approximately 85\% of the total sample. The two key observables used in this study are the uncorrected PAH flux measurements in Table 4 of \citet{Lai2020} and the silicate optical depths (\tauSi), which can be inferred from the reported silicate strength ($S_{Sil}$) in \citet{Spoon2022} using Equation~\ref{equ:tauSi}.

\begin{figure}
  \centering
  \includegraphics[width=\columnwidth]{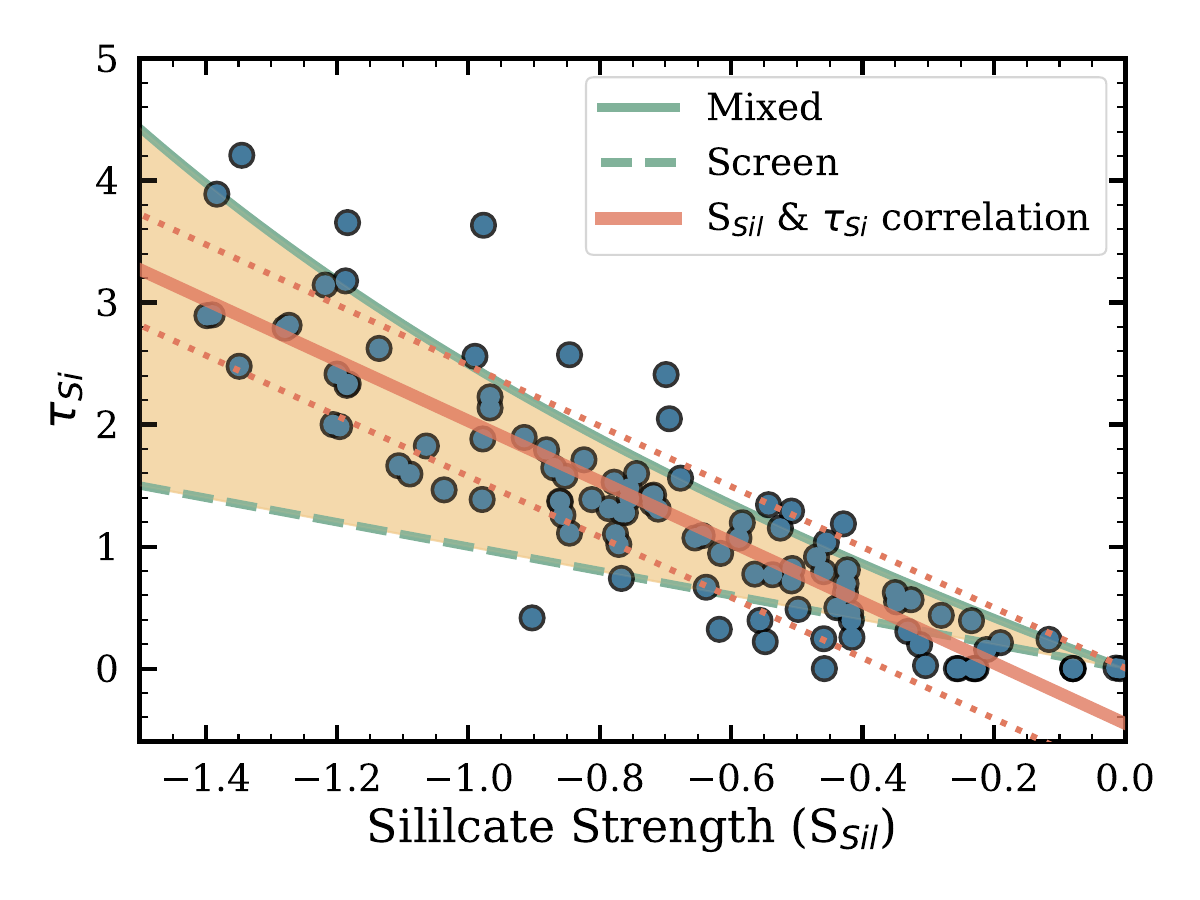}
  \caption{An empirical relationship between the PAHFIT fitted \tauSi\ and the silicate strength (S$_{Sil}$) taken from IDEOS. The red line is the linear fit to these two parameters with the dotted indicating the 1-$\sigma$ range. Overall, the points lie within the shaded region bracketed by the mixed and screen geometries.}
  \label{Fig:tauSi_and_SilStrength}
\end{figure}

The wavelength at 9.7\um, where the MIR attenuation peaks, is often used for constraining the level of the attenuation in a galaxy. The amount of how much the continuum at this wavelength is extinguished depends on both \tauSi\ and the adopted attenuation curve (see Equation~\ref{eq: att} for detail). As in this study, our aim is to constrain the attenuation curve, an independent measurement of the silicate optical depth is needed. This is why instead of using the PAHFIT fitted \tauSi\ values, which were derived based on a mixed geometry together with the prescribed attenuation curve, we obtained the silicate opacity based on a distinct measurement of the silicate strength reported in \citet{Spoon2022}\footnote{The silicate strength is defined as $S_{Sil}=ln\frac{f_{\nu}(9.7\um)}{f_{\nu}^{cont}(9.7\um)}$, where $f_{\nu}(9.7)$ is the flux density of the spectrum at the peak of the silicate absorption, and $f_{\nu}^{cont}(\lambda_{peak})$ is the flux density of the underlying continuum at the same wavelength estimated by the spline method.}. To convert S$_{Sil}$ to \tauSi, we invoke the empirical correlation based on the \citet{Lai2020} bright PAH sample (Figure~\ref{Fig:tauSi_and_SilStrength}), which can be characterized as:

\begin{equation}
    \label{equ:tauSi}
    \tauSi = -2.48 \times S_{Sil} - 0.45
\end{equation}

\noindent provided that \tauSi$\geq$0. This relationship mainly lies within the range between the mixed and screen geometries and has a 1-$\sigma$ deviation of 0.45. Hereafter, when referring to \tauSi, it points to the silicate optical depth that is derived from S$_{Sil}$.

\section{Constraining the MIR attenuation curve using PAH bands}
\label{sec:model}

In \citet{Lai2020}, two geometries were explored in the modeling for the MIR spectral decomposition --- fully mixed and obscured continuum geometries, bracketing realistic scenarios of PAH emission and silicate absorption in galaxies. We recall that the obscured continuum scheme represents a geometry consisting of a concentrated continuum emitting source plus unobscured PAH emitters atop the continuum. In such a case, PAH band strengths are measured with their local continua, and their emission sources are presumed to lie in the ``foreground" of the extinction-impact continuum sources. Thus, PAH emission remain free from extinction correction.  

\begin{figure*}
  \centering
  \includegraphics[width=0.8\textwidth]{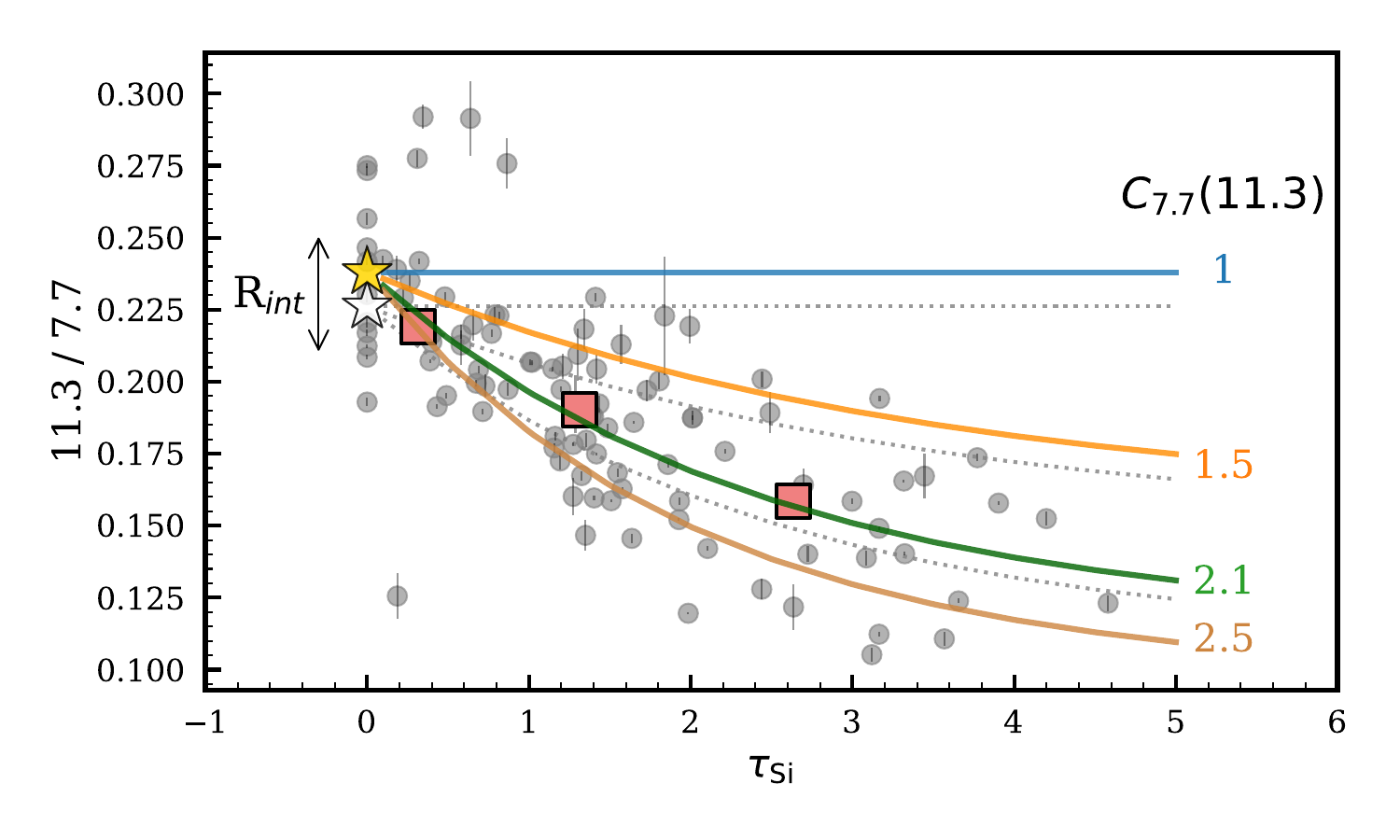}
  \caption{A schematic view of our PAH attenuation model. The data (circles) represent the correlation between the uncorrected PAH band flux ratio 11.3/7.7 and the continuum-determined silicate optical depth (\tauSi). The squares indicate values that are binned with an equal number of galaxies. Our model uses two parameters --- $C_{7.7}(\lambda)$ and \Rint\ --- to characterize the distribution. The term $\lambda$ is the central wavelength of the selected PAH band with an index of $i$. In this case, $C_{7.7}(11.3)$ represents the relative attenuation at 11.3\um\ with respect to that at 7.7\um\ (see Section~\ref{sec: Grid_model} for details). It is unity if there is no differential attenuation between 11.3 and 7.7\um; however, in cases where there is more attenuation at 11.3\um\ compared to 7.7\um, the curves exhibit steeper downward trends as \tauSi\ increases. The value of $R_\mathrm{int}$, denoted by the star, represents the intrinsic, unattenuated band ratio of 11.3/7.7 that anchors the starting point of each $C_{7.7}(11.3)$ curve. Here we plot two families of curves for illustration. The dotted curves are shifted arbitrarily lower than the solid curves by 5\%, which is controlled by the $R_\mathrm{int}$ value. The $C_{7.7}(11.3)$ values listed next to their corresponding curves control the strength of the band ratio variation with \tauSi. The best fit to the distribution of the observation is $R_\mathrm{int}$=$0.238^{+0.006}_{-0.006}$ and $C_{7.7}(11.3)$=$2.072^{+0.366}_{-0.159}$.}
  \label{Fig:grid_model}
\end{figure*}

Whether or not the assumption of pure obscured continuum geometry holds in galaxies can be verified by testing if the silicate optical depth (\tauSi) and PAH band ratios correlate. In other words, if the geometry were to be obscured continuum, meaning that there is no need for extinction correction for the measured PAH band strengths, band ratios ought to exhibit no trend with respect to \tauSi. Otherwise, if a trend exists, it is likely that assuming all galaxies have PAH emission atop a continuum produced by central obscured sources is incorrect. Taking PAH band ratio 11.3/7.7 for an example, in Figure~\ref{Fig:grid_model}, we show the relationship between \tauSi\ and the 11.3/7.7 ratio measured in the obscured continuum geometry. It is apparent that there is a decreasing trend of 11.3/7.7 as \tauSi\ increases, suggesting the assumption of an obscured continuum geometry that accounts for all galaxies is inappropriate, though it may be applicable to a subset of galaxies, and can still be regarded as a useful physical limit. This decreasing trend indicates that PAH emissions at 11.3 and 7.7\um\ are indeed subjected to differential extinction. 

Conventionally, the most widely used approach to determine dust extinction is the Balmer decrement (e.g., H$\alpha$/H$\beta$). The advantage of using recombination lines to study the influence of dust absorption is that the ratio is largely fixed by quantum physics, and any deviation of the observed line ratio from the theoretical value (e.g., case B condition in \citet{Storey1995}) can be attributed to extinction. We follow a methodology conceptually similar to the pioneering use of the Balmer decrement to determine dust extinction \citep{Kennicutt1992, Dominguez2013} by attempting to constrain the shape of the MIR extinction curve using major PAH bands at 3.3, 6.2, 8.6, 7.7, 12.6, and 17\um. Although there is no a priori theoretical expectation of the PAH band luminosity ratios, as long as there is a trend (even with scatter) between \tauSi\ and the uncorrected PAH band ratio, we can infer an intrinsic, unattenuated value for the band ratio and attribute deviations in the ratio with varying \tauSi\ to the differential extinction. \citet{Hernan-Caballero2020} used a similar concept to measure \tauSi\ in ULIRGs; specifically, these authors employed the fact that the intrinsic PAH 12.6/11.3 ratio remains relatively constant (with variations $\lesssim$5\%) and that the 11.3\um\ PAH is more susceptible to attenuation compared to the 12.6\um\ PAH, as the former sits closer to the 9.7\um\ silicate absorption band. 

The differential attenuation between the selected PAH bands can thus be useful to constrain the shape of the MIR attenuation curve. In this regard, we are able to investigate the shape of the MIR attenuation curve by flipping the problem from assuming a ``perfect" attenuation curve and then measuring the PAH strengths to using uncorrected PAH band strengths to ``constrain" the shape of the attenuation curve itself. The bright PAH galaxies in ASESS \citep{Lai2020} constitute a good test for this attenuation investigation because they cover all the PAH bands from 3--18\um, in particular, enabling us to study the attenuation slope from 3 to 7\um\ and the strength of the 18\um\ silicate feature.



\subsection{The PAH Attenuation Model}
\label{sec: Grid_model}
We selected major PAH bands at 3.3, 6.2, 7.7, 8.6, 11.3, 12.6, and 17\um\ to constrain the shape of the MIR attenuation curve. Based on our bright PAH sample, we found $L_\mathrm{PAH 7.7}/\LIR$ has the smallest \emph{normalized} standard deviation, implying the strength of PAH 7.7 is relatively stable compared to the integrated IR luminosity. The 7---8\um\ range is also at a relatively low-point in current extinction curves, falling between the rising power-law extinction at shorter wavelengths, and the strong silicate resonance at 9.7\um.  We thus choose 7.7\um\ as a fiducial point and use ratios with respect to this point to measure relative attenuation at different wavelengths, which can be expressed as $C_{7.7}(\lambda)$ as mentioned in Section \ref{sec:extinction_curve_comparison}. 

In the PAH attenuation model, we attempt to constrain the differential attenuation between two PAH bands by assuming the PAH emitter and the dust absorber are well mixed. In such a mixed geometry, the emission is subjected to the extinction of a factor of $(1-e^{-\tau_{\lambda}})/\tau_{\lambda}$. The term $\tau_{\lambda}$ features the wavelength-dependent opacity and can be written as

\begin{equation}
    \tau_{\lambda} = \tau_\mathrm{Si} \cdot C_{9.7}(\lambda)
    \label{eq: att}
\end{equation}

\noindent with $\tau_\mathrm{Si}$ denoting the silicate optical depth measured at 9.7\um\ and $C_{9.7}(\lambda)$ the extinction curve normalized at 9.7\um. To investigate how extinction can alter PAH band ratios, we first assume the intrinsic PAH flux ratio (\Rint) between a particular PAH band at wavelength $\lambda_i$ and the 7.7\um\ PAH to be

\begin{equation}
    \Rint = \frac{F_{\lambda_i}}{F_{7.7}},
\end{equation}

\noindent In the mixed geometry, the observed attenuated ratio between a given PAH band and the 7.7\um\ PAH (\Robs) can be characterized as:

\begin{equation}
  \Robs = \frac{F_{\lambda_i}}{F_{7.7}} \frac{\left( 1-e^{-\tau_{\lambda_i}} \right)}{\left( 1-e^{-\tau_{7.7}} \right)}\frac{\tau_{7.7}}{\tau_{\lambda_i}}
\end{equation}

\noindent Together, Equations~2--4 give the ratio between \Robs\ and \Rint\ as

\begin{equation}
  \frac{\Robs}{\Rint} = \frac{\left( 1-e^{-\tauSi\, C_{9.7}(\lambda_i)} \right)}{\left( 1-e^{-\tauSi\, C_{9.7}(7.7)} \right)}\frac{C_{9.7}(7.7)}{C_{9.7}(\lambda_i)}
\end{equation}

\noindent Noting that $C_{9.7}(7.7)/C_{9.7}(\lambda_i)=1/C_{7.7}(\lambda_i)$ yields an attenuation model 

\begin{equation}
  \frac{\Robs}{\Rint} = \frac{\left( 1-e^{-\tauSi \cdot C_{9.7}(7.7) \cdot C_{7
  .7}(\lambda_i)} \right)}{\left( 1-e^{-\tauSi\, C_{9.7}(7.7)} \right)}\frac{1}{C_{7.7}(\lambda_i)}
\end{equation}

\noindent The attenuation curve constraint problem can thus be expressed in a Bayesian form as:

\begin{multline}
    P(\Rint, C_{7.7}(\lambda_i) | \Robs, \tauSi, C_{9.7}(7.7)) \propto \\
    P(\Robs, \tauSi, C_{9.7}(7.7)| \Rint, C_{7.7}(\lambda_i)) \cdot P(\Rint, C_{7.7}(\lambda_i))
\end{multline}

\noindent such that \Rint\ and $C_{7.7}(\lambda)$ are the posteriors we infer given observations of \Robs, \tauSi, and $C_{9.7}(7.7)$.  In Figure~\ref{Fig:grid_model}, we use the 11.3/7.7 PAH band ratio to demonstrate the concept of using \Rint\ and $C_{7.7}(\lambda)$ to fit the observables. If there were no differential extinction between 11.3 and 7.7\um, implying $C_{7.7}(11.3)=1$, the observed 11.3/7.7 ratios would be expected to be distributed horizontally, independent of \tauSi. As $C_{7.7}(11.3)$ increases, the predicted distribution curves down towards higher \tauSi. The locus at \tauSi=0, serving as the common origin for various $C_{7.7}(11.3)$ curves, signifies the intrinsic PAH band ratio, \Rint. Treating \Rint\ as a free parameter gains leverage in fitting the data because we have limited a priori knowledge of the intrinsic PAH band ratios in the sample. A comparison between the estimated \Rint\ and the values derived from the nearby SINGS galaxies will be discussed in Section~\ref{sec:Discussion}. 

\begin{figure*}
  \centering
    \includegraphics[width=\textwidth]{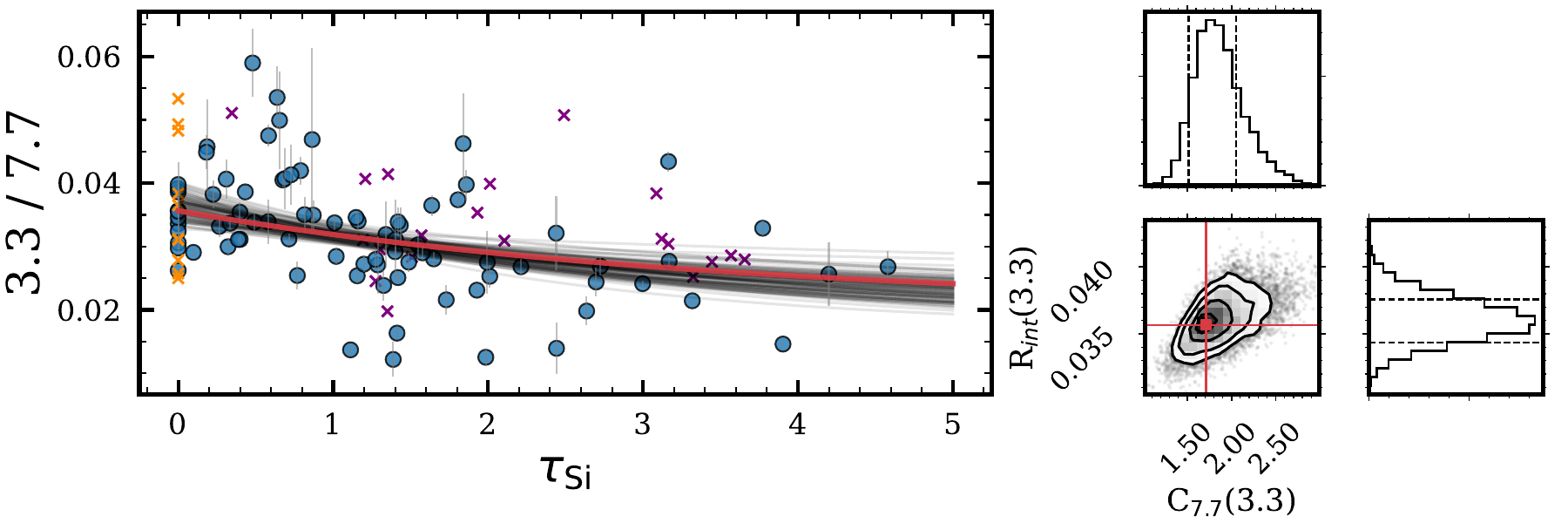}
    \includegraphics[width=\textwidth]{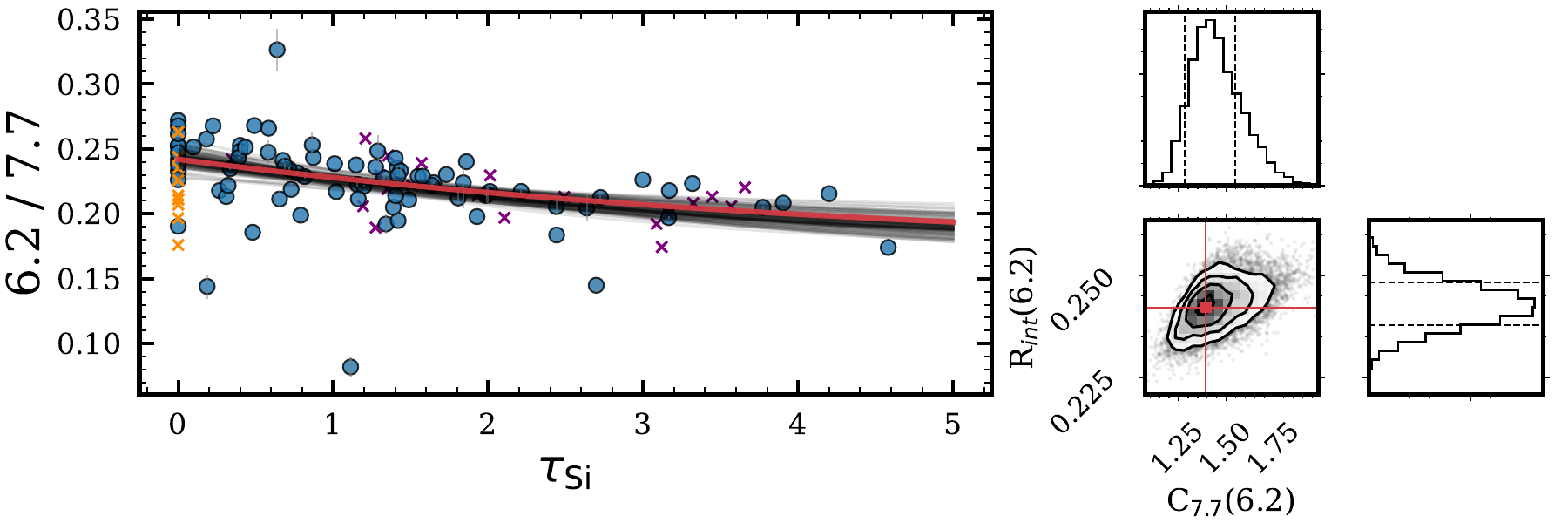}
    \includegraphics[width=\textwidth]{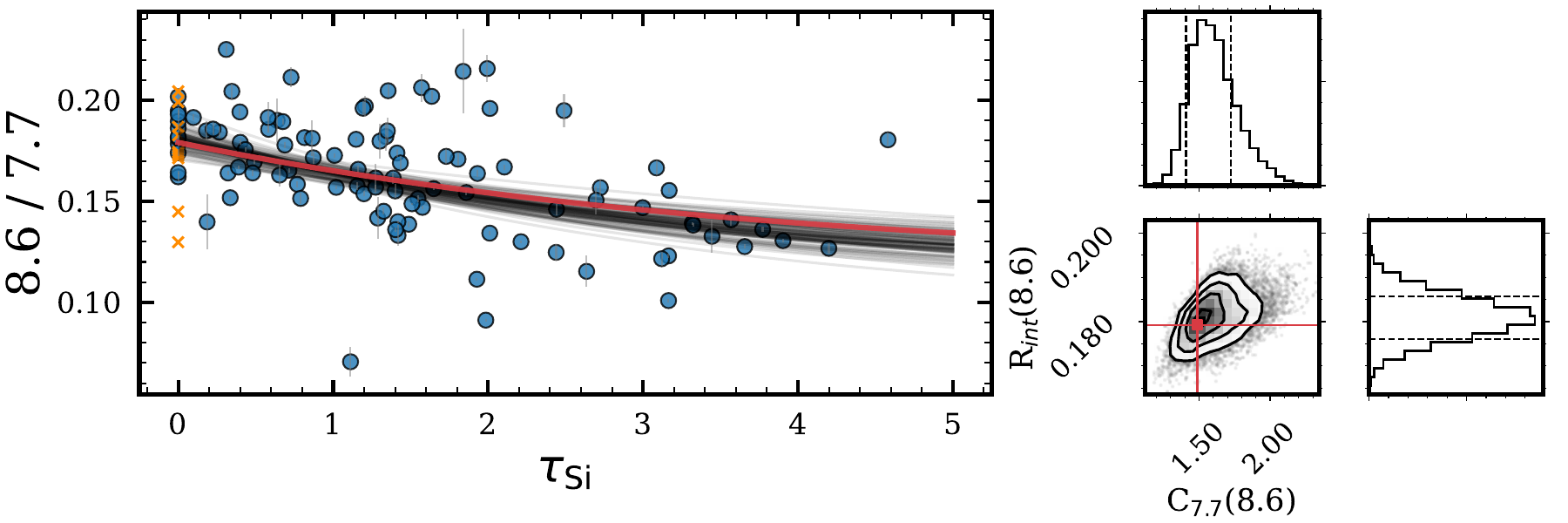}
      \caption{\textit{Left:} Correlation between each uncorrected PAH band ratio and \tauSi, with 100 randomly selected samples from the chain of MCMC. The red line indicates the curve that is characterized by the median value of the posterior distribution. Cross symbols are galaxies excluded from our analysis (see Section~\ref{sec: Results} for detail). \textit{Right:} Corner plot of the posterior distributions of $C_{7.7}(\lambda)$ (upper sub-panels) and \Rint\ (right sub-panels). The red point indicates the values of $C_{7.7}(\lambda)$ and \Rint\ used in plotting the red curve on the left panel. The 16$^{th}$ and 84$^{th}$ percentiles of the marginalized distributions are indicated in dashed lines.}
  \label{Fig:MCMCfit_Rint_Rtau}
\end{figure*}

\begin{figure*}
  \setcounter{figure}{3}
  \centering
    \includegraphics[width=\textwidth]{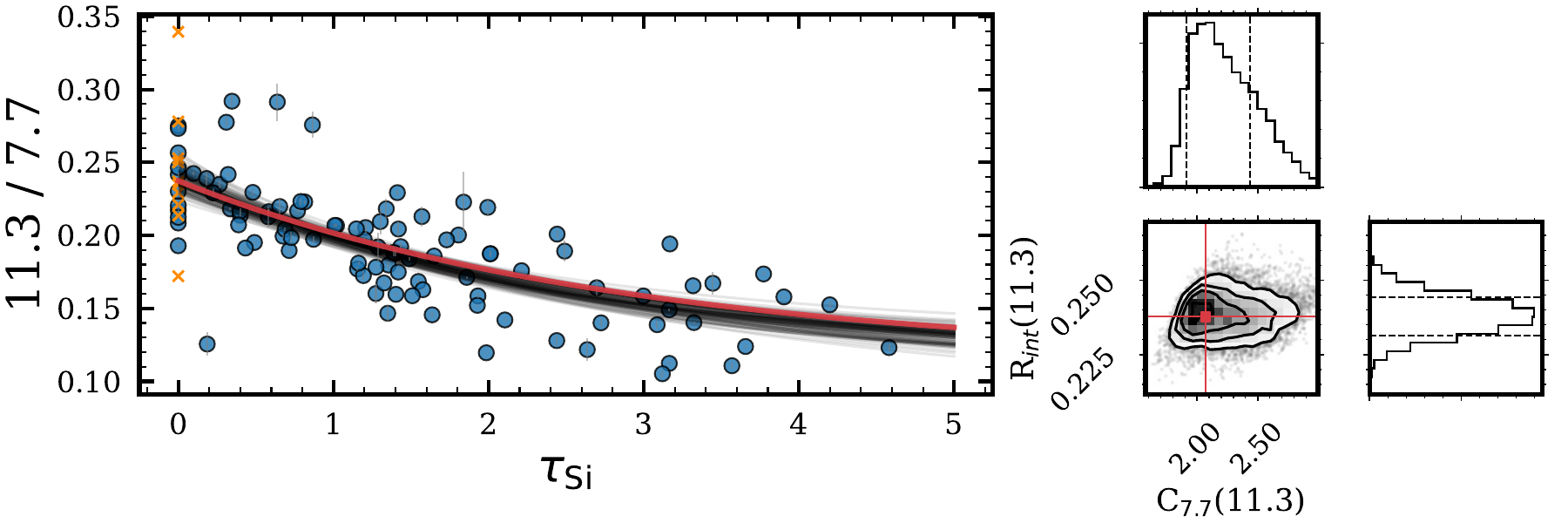}
    \includegraphics[width=\textwidth]{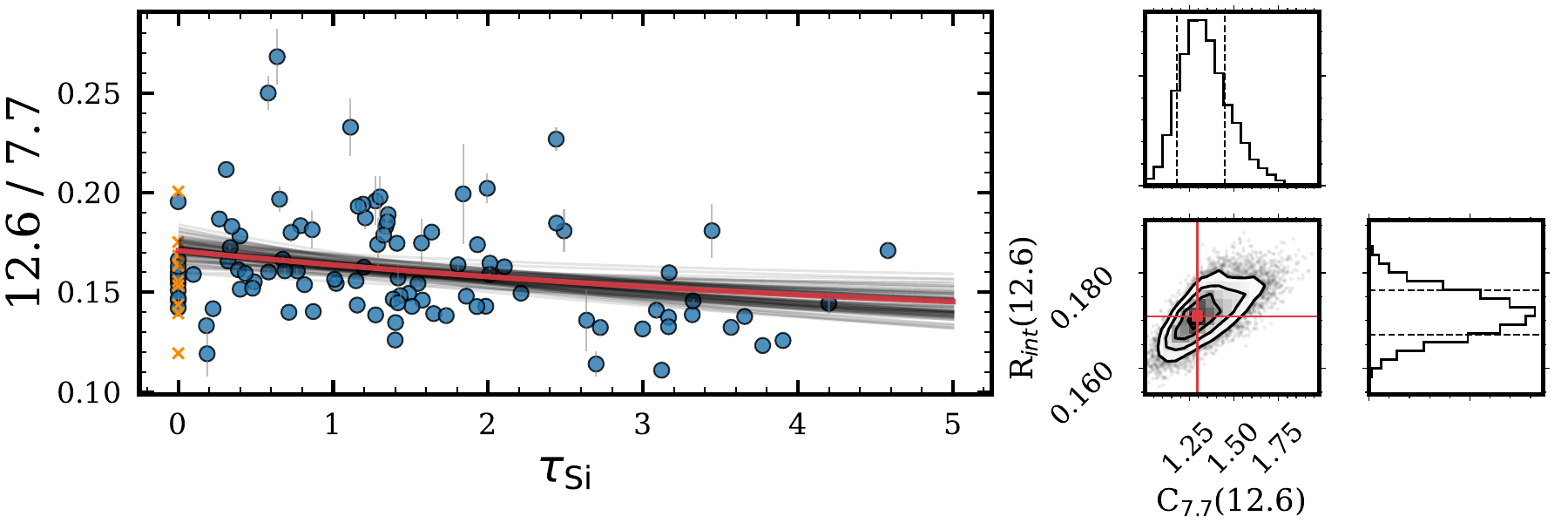}
    \includegraphics[width=\textwidth]{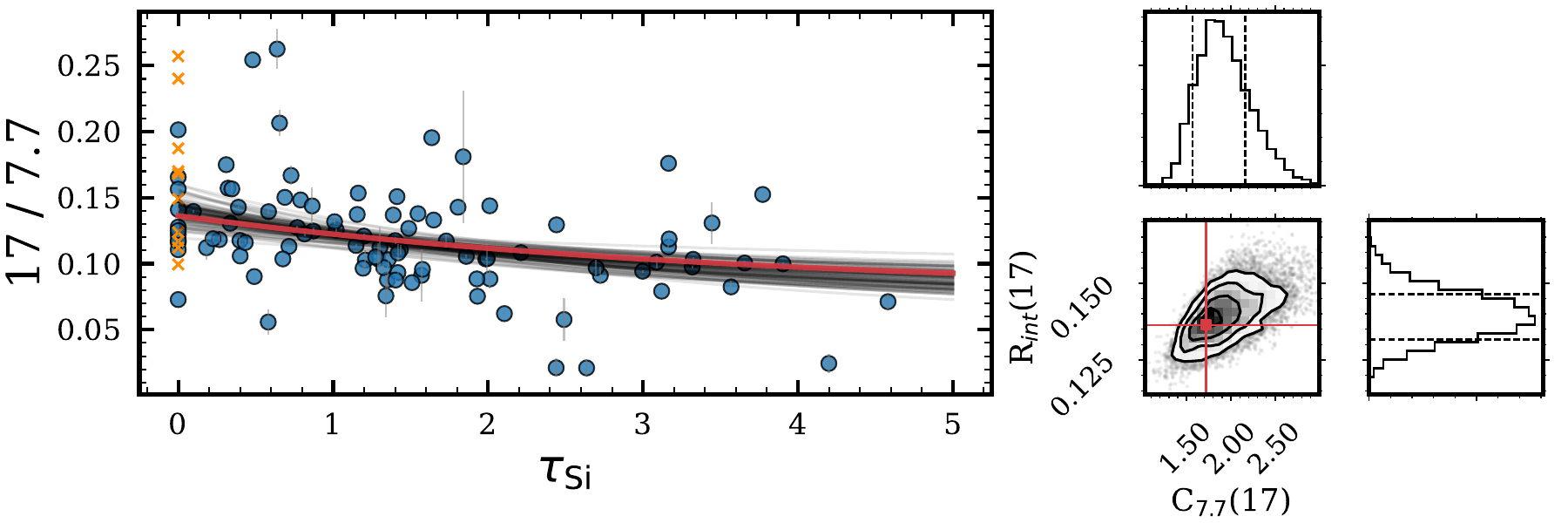}
  \caption{(continued.)}
  
\end{figure*}

\subsection{Differential Attenuation among PAH Bands}
\label{sec: Results}
With this model, we are able to use PAH emissions to estimate the differential attenuation between 7.7\um\ and the wavelength positions of other bright PAH bands, including 3.3, 6.2, 8.6, 11.3, 12.6, and 17\um. In Figure~\ref{Fig:MCMCfit_Rint_Rtau}, we show the trend in the ratios of different PAH band pairs against \tauSi\ measurements. Uncorrected PAH measurements are taken from the ASESS bright PAH sample listed in Table 4 of \citet{Lai2020}. Note that \PAHFIT\ has limited sensitivity to opacities \tauSi $\leq$ 0.2 \citep{Smith2007}, so galaxies exhibiting low \tauSi\ span a wider range in band-ratio space. PAH band strengths are likely not that well constrained in galaxies with possible AGN contribution, which may lead to a flatter continuum slope beyond 15\um\ \citep{Laurent2000, Brandl2006}. Therefore, we refine our sample by excluding those galaxies that have \tauSi$<$0.01 and also exhibit flatter continuum slopes ($f_{\nu}(30)/f_{\nu}(15)<$6) from the analysis (removed galaxies are shown in orange crosses in the figure). Only 10 galaxies are removed from this cut ($<$10\% of 113 galaxies in the ASESS bright PAH sample). In addition, galaxies with high \tauWater ($\geq$0.4) at 3\um\ were removed when studying the PAH 3.3/7.7 and 6.2/7.7 ratios because water absorptions at 3 and 6\um\ may impact the 3.3 and 6.2\um\ PAH flux measurements, although the impact can be limited as discussed in Section~\ref{sec:Caveats}. The 20 galaxies that are removed with this criterion are shown in purple crosses. 

The \Rint\ and $C_{7.7}(\lambda)$ posterior contours are derived by employing a Markov chain Monte Carlo (MCMC) analysis, implemented using $emcee$ \citep{Foreman-Mackey2013}. We generate 5,000 sampling steps of MCMC with 500 burn-in, assuming flat priors on \Rint\ and $C_{7.7}(\lambda)$. The variable $C_{9.7}(7.7)$ is treated as a nuisance parameter and is assigned a flat prior from 0.2---0.5, bracketing the range of values found in the S07 and W15 curves (see Figure~\ref{fig:extinction_curve_comparison}(a)). The red line in Figure~\ref{Fig:MCMCfit_Rint_Rtau} shows the resulting fit for each PAH band ratio combination. The left panel includes the data and the curves generated from the 100 randomly selected posteriors from the MCMC chain. The right panel shows the corner plot of the two posterior distributions of \Rint\ and $C_{7.7}(\lambda)$. The fitted results are the median of the posterior distributions, with associated uncertainties based on the 16$^{th}$ and 84$^{th}$ percentiles of the marginalized distributions (see Table~\ref{tab:Rint_and_Rtau}). For reference, the SINGS $R_\mathrm{int}$ values taken from \citet{Smith2007} are also included in the table. Clearly, the MCMC-fitted distributions are far from horizontal, but with curvatures in different degrees, with the most differential attenuation between 7.7 and 11.3\um\ ($C_{7.7}(11.3)=2.072$) and the least between 7.7 and 12.6\um\ ($C_{7.7}(12.6)=1.294$). This result implies the assumption of an obscured continuum geometry clearly does not apply to a large majority of sources. PAH bands are unambiguously subjected to attenuation.

\begin{figure*}
  \centering
  \includegraphics[width=\textwidth]{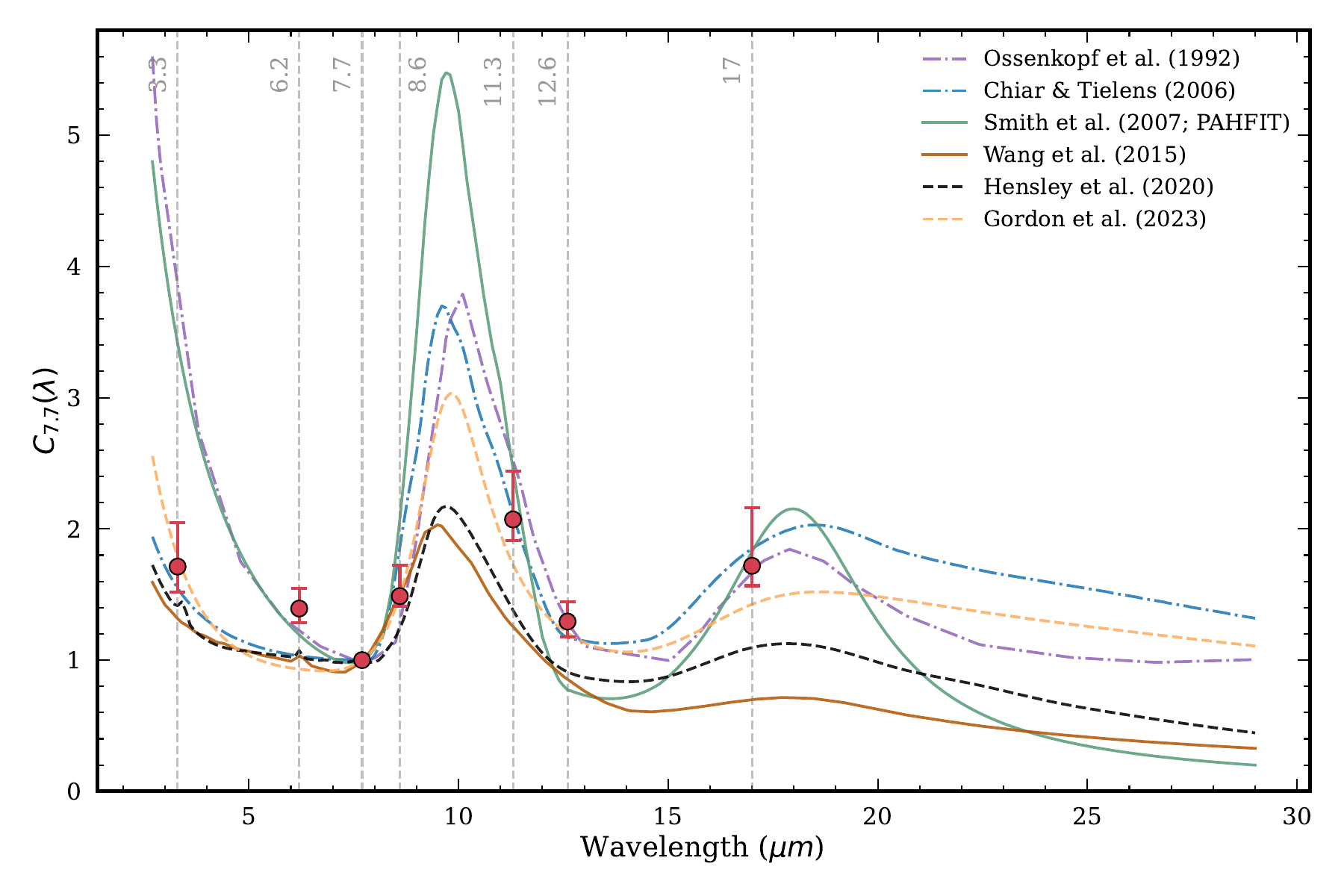}
  \caption{PAH-ratio based constraints on the aggregate MIR attenuation curve of galaxies measured in this study. Pre-existing literature extinction curves are shown as solid colored lines, normalized at 7.7\um. The red points represent the measured differential attenuation values (relative to 7.7\um) recovered by our model at 3.3, 6.2, 8.6, 11.3, 12.6, and 17\um\, with error bars indicating the ranges of 16$^{th}$ and 84$^{th}$ percentiles of the posterior distribution of $C_{7.7}(\lambda)$.
  }
  \label{Fig:extinction_curve_constraints_PAH}
\end{figure*}

Finally, as we overlay our $C_{7.7}(\lambda)$ measurements on the extinction curves presented in Figure~\ref{fig:extinction_curve_comparison}, they generally lie within the envelope of these extinction curves (see Figure~\ref{Fig:extinction_curve_constraints_PAH}). This shows that the uncorrected PAH band ratios are dependent on \tauSi\ in a manner that is consistent with a range of reasonable extinction curves. Overall, our result agrees more with CT06. At shorter wavelengths ($\lambda<8$\um), the trend in band ratios is more consistent with ``flatter'' extinction curves, which in turn shows that the strong power-law rise in extinction from 8 to 3\um\ as assumed by \PAHFIT\ is not well supported. In the wavelength range near the peak of the 9.7\um\ feature (8\um$\leq \lambda<$13\um), our result shows a better match to O92, CT06, and G23 in terms of the width of the silicate feature. At longer wavelengths, though not well constrained, the PAH ratio constraint near the 18\um\ absorption band is more consistent with extinction models that exhibit a strong and/or narrower secondary silicate O-Si-O resonance such as O92, CT06, and S07.

\begin{deluxetable}{cccc}
\label{tab:Rint_and_Rtau}
\tabletypesize{\footnotesize}
\tablewidth{0pt}
 \tablecaption{Fitted result of $R_{int}$ and $C_{7.7}(\lambda)$}

\tablehead{
 \colhead{PAH band } & \colhead{$R_{\mathrm{int},i}$} & \colhead{$C_{7.7}(\lambda)$} & \colhead{SINGS $R_\mathrm{int}$} \\ [-0.2cm]
  \colhead{} & \colhead{} & \colhead{} & \colhead{(10\%, 50\%, 90\%)} \\ [-0.4cm]
 }
 \startdata 
 3.3\um\ & $0.036^{+0.002}_{-0.002}$ & $1.713^{+0.335}_{-0.196}$ & --- \\ 
 6.2\um\ & $0.243^{+0.005}_{-0.005}$ & $1.393^{+0.153}_{-0.110}$ & 0.21, 0.28, 0.75 \\
 8.6\um\ & $0.181^{+0.005}_{-0.005}$ & $1.488^{+0.236}_{-0.079}$ & 0.11, 0.18, 0.21 \\
 11.3\um\ & $0.238^{+0.006}_{-0.006}$ & $2.072^{+0.366}_{-0.159}$ & 0.21, 0.28, 0.67 \\
 12.6\um\ & $0.171^{+0.005}_{-0.005}$ & $1.294^{+0.152}_{-0.117}$ & 0.12, 0.16, 0.27\\
 17.0\um\ & $0.140^{+0.008}_{-0.007}$ & $1.719^{+0.444}_{-0.152}$ & 0.09, 0.15, 0.32 \\
 \enddata

 \tablecomments{The result of \Rint\ and $C_{7.7}(\lambda)$ in different PAH badns. The term SINGS $R_\mathrm{int}$ refers to the PAH band ratios observed in SINGS galaxies, as reported by \citet{Smith2007}. These ratios are presented with their respective 10\%, 50\%, and 90\% percentile ranges.}
\end{deluxetable}

\section{Caveats}
\label{sec:Caveats}
Several caveats are worth mentioning in our analysis. Although we conclude that obscured continuum geometry cannot account for the strong variation in PAH band ratio with continuum opacity, this statement is based on the \citet{Lai2020} ``bright PAH'' galaxies,  the majority of which are is comprised of dusty, metal-rich star-forming (U)LIRGs, with only moderate AGN contribution. The reason that the obscured geometry was invoked in the first place was to address the obscuration conditions in galaxies powered by a buried AGN or dominant compact central starburst. So, in that sense, it is not surprising that our conclusions disfavor obscured geometry and lean more towards a fully mixed scenario. 

It is also worth discussing screen geometry, in which all emitters are situated behind a layer of dust that is responsible for all observed extinction. We have tested our attenuation model against an alternative assumption of screen geometry and found the model's outcomes are robust between the choice of the screen and mixed geometry. In the screen geometry, the differential attenuation between two bands can reach a given level with a smaller \tauSi\ value, since all sources are subjected to the same level of screen obscuration, as illustrated in Figure~\ref{Fig:tauSi_and_SilStrength}. For instance, a mixed \tauSi=3 in the empirical relationship yields a silicate strength comparable to that from \tauSi=1.4 in screen geometry. By fully replicating the methods of Section~\ref{sec: Results} and incorporating the necessary adjustments for the input \tauSi\ values relevant to screen geometry, we find the estimated C$_{7.7}(\lambda$) values change very little, increasing by 7.6\% on average --- well within the uncertainties.






We attribute the change of PAH ratios with respect to \tauSi\ solely to differential attenuation in our analysis. Yet, there might be other second-order dependencies of the intrinsic PAH ratios, such as metallicity and AGN presence, which are themselves correlated with \tauSi. These may explain some of the scatter in Figure~\ref{Fig:MCMCfit_Rint_Rtau}, although we attempted to control for these by careful sample selection, focusing on dusty (U)LIRGs primarily powered by starburst. Note that these second-order effects can be mitigated as our band ratios are obtained from galaxy-integrated values. Also, we note that crystalline silicate in deeply obscured galaxies shapes the extinction curve, which, in particular, may affect the measurements of PAH bands at 11.3 and 17\um\ \citep{Spoon2006, Spoon2022}. However, given that our sample mainly exhibits low to moderate \tauSi, with an average of S$_{Sil}$=-0.7 as opposed to those galaxies with S$_{Sil}<-1$ that exhibits appreciable crystalline silicate absorption \citep{Spoon2022}, we expect little or no impact from crystallinity.

Although our result suggests that the attenuation curve is relatively flat in the region between 3--8\um, the impact of ice absorption centered at 3.05\um\ was not taken into account in the analysis, and this absorption feature, if strongly present, can certainly impact the 3.3\um\ PAH measurement. To understand at what level \tauWater\ affects our result, we follow the MCMC analysis but only apply it to the low \tauWater\ sample, with \tauWater$\leq$0.2, the median of \tauWater\ in the bright PAH sample; the $C_{7.7}(3.3)$ value inferred from this low \tauWater\ sample changes by only 9\% ($C_{7.7}(3.3)$=1.563) compared to the full sample estimate, well within the uncertainty. More investigation is needed to carefully separate the absorption effects of ice from the NIR slope at the $\sim$3\um\ regime in highly obscured sources. 

\section{DISCUSSION}
\label{sec:Discussion}
As indicated in Section~\ref{sec:sample}, our sample predominantly consists of (U)LIRGs. The evident correlation between the silicate optical depth and PAH ratios, as shown in Figure~\ref{Fig:MCMCfit_Rint_Rtau}, implies that the majority of galaxies in our sample cannot be explained solely by the obscured continuum geometry presented in \citet{Lai2020}. Nevertheless, this observation does not eliminate the possibility of such geometry existing in a limited subset of individual galaxies, or in greater frequency among more obscured systems than considered.

The attenuation curve constraints presented in Section~\ref{sec: Results} lie broadly within the range of currently known Milky Way extinction curves. At wavelengths shorter than 8\um, our result suggests an intermediate slope, slightly favoring a flatter (grayer) curve, but slightly steeper than what has been proposed in CT06, W15, and H20.  Previous studies have attributed this flattening to dust grain growth through grain coagulation or ice mantle coating, as the shallower slope can be explained by scattering from larger grains \citep[e.g.][]{Flaherty2007, Chapman2009, McClure2009}.  Measurements between 8---13\um\ allow us to constrain the width of the silicate feature, but most importantly, as shown in Figure~\ref{Fig:extinction_curve_constraints_PAH}, the variation of the width depends largely on the differential attenuation between the local extinction minimum at 7.7\um\ and the peak at 9.7\um\ given the way the curves are normalized.  Our result suggests a better match with curves of O92, CT06, and G23 in the vicinity of the silicate feature, predominantly around the  $3.0\leq C_{7.7}(9.7)\leq4.0$ range.  Given our constraints at 11.3 and 12.6\um, our findings agrees well with those presented in \citet{Hernan-Caballero2020}, indicating that the CT06 profile provides a reasonable representation of the attenuation level in 11---13\um. Finally, at longer wavelengths, our measurement at 17\um\ is in line with those curves that show a stronger 18\um\ silicate profile. We find a strong 18\um\ bump, such as the \PAHFIT\ adopted S07, is necessary for MIR spectra decomposition.

Aside from constraining the attenuation at wavelengths where PAH bands are located, our model also provides estimates of the intrinsic, unattenuated PAH band ratios (R$_{int}$) in luminous star-forming systems (see Table~\ref{tab:Rint_and_Rtau}). These ratios agree well with the values derived from low-attenuation SINGS galaxies reported in \citet{Smith2007}, typically differing by only 5\%, with 11.3/7.7 ratio having the largest difference (15\%). 

Our method of using PAHs to constrain MIR attenuation would greatly benefit from measurements near the peak of the silicate absorption resonances at 9.7 and 18\um; unfortunately, there are no available strong PAH features in these regions. Nonetheless, emission features other than PAHs, such as the \molH\ pure rotational lines, can be used in a related analysis, and will be addressed in \paperII.

\section{Summary}
We introduce a novel approach using the full suite of PAH emission bands from the large combined AKARI+Spitzer ASESS sample to constrain the attenuation curve of galaxies in the MIR range.  With the simple assumption that PAH bands have relatively fixed intrinsic ratios among metal-rich star-forming galaxies, we investigate the differential attenuation at PAH-specific wavelengths to constrain the \emph{shape} of the attenuation curve. In particular, our focus is on constraining the MIR attenuation curve at six distinct wavelengths where prominent PAH features are located (3.3, 6.2, 8.6, 11.3, 12.6, and 17\um). To our knowledge, this is the first time the detailed shape of the aggregate MIR attenuation curve in a large sample of galaxies has been determined empirically. Our major findings are summarized as follows:

\begin{itemize}
    \item We find PAH features are indeed subject to differential attenuation as shown in the downward trends between the ratios of PAHs and the silicate optical depth (\tauSi).
    
    \item The attenuation profile we measured generally lies within the spread of the pre-existing extinction curves, favoring curves with a grayer attenuation between 3---8\um\ as well as those with relatively high levels of attenuation at 17\um\ caused by the silicate O-Si-O resonance. The derived profile is independent of the assumed attenuation geometry.
    
\end{itemize}

As the most prominent absorption feature from far-UV to FIR wavelengths, the silicate 9.7\um\ resonance plays an important role in measuring obscuration in galaxies, with a strong impact on MIR spectral decomposition. While we can derive the attenuation level at certain wavelengths across 3--17\um\ using the available PAH features, the amount of attenuation at 9.7\um\ cannot be directly obtained, as no PAH band resides near the absorption peak.  Fortuitously, there is one \molH\ pure rotational line sitting directly at the peak of the 9.7\um\ silicate absorption feature, allowing us to investigate the possibility and power of using the suite of \molH\ emissions to further constrain the attenuation curve at this crucial wavelength. An in-depth study using \molH\ to provide further constraints on the shape of the MIR attenuation curve will be detailed in \paperII.


\begin{acknowledgments}
The authors thank A. Witt for valuable discussions. TSYL extends special thanks for the hospitality provided by JAXA and acknowledges funding support from NASA grant JWST-ERS-01328. JDTS gratefully acknowledges support for this project from the Research Corporation for Science Advancement through Cottrell SEED Award No. 27852. EP acknowledges support from the Natural Sciences and Engineering Research Council of Canada. MI is supported by JP21K03632. TN acknowledges the support by JSPS KAKENHI Grants Number 21H04496 and 23H05441.

\vspace{5mm}
\facilities{AKARI, Spitzer}


\software{Astropy \citep{Astropy2013, Astropy2018},
          Bokeh \citep{Bokeh2018},
          Matplotlib \citep{Hunter2007},
          Numpy \citep{VanderWalt2011},
          \PAHFIT\ \citep{Smith2007, Lai2020},
          SciPy \citep{Virtanen2020}
          }

\end{acknowledgments}


\bibliography{Lai_attenuation}

\begin{thebibliography}{}
\expandafter\ifx\csname natexlab\endcsname\relax\def\natexlab#1{#1}\fi
\providecommand{\url}[1]{\href{#1}{#1}}
\providecommand{\dodoi}[1]{doi:~\href{http://doi.org/#1}{\nolinkurl{#1}}}
\providecommand{\doeprint}[1]{\href{http://ascl.net/#1}{\nolinkurl{http://ascl.net/#1}}}
\providecommand{\doarXiv}[1]{\href{https://arxiv.org/abs/#1}{\nolinkurl{https://arxiv.org/abs/#1}}}

\bibitem[{{Armus} {et~al.}(2004){Armus}, {Charmandaris}, {Spoon}, {Houck}, {Soifer}, {Brandl}, {Appleton}, {Teplitz}, {Higdon}, {Weedman}, {Devost}, {Morris}, {Uchida}, {van Cleve}, {Barry}, {Sloan}, {Grillmair}, {Burgdorf}, {Fajardo-Acosta}, {Ingalls}, {Higdon}, {Hao}, {Bernard-Salas}, {Herter}, {Troeltzsch}, {Unruh}, \& {Winghart}}]{Armus2004}
{Armus}, L., {Charmandaris}, V., {Spoon}, H.~W.~W., {et~al.} 2004, \apjs, 154, 178, \dodoi{10.1086/422915}

\bibitem[{{Astropy Collaboration} {et~al.}(2013){Astropy Collaboration}, {Robitaille}, {Tollerud}, {Greenfield}, {Droettboom}, {Bray}, {Aldcroft}, {Davis}, {Ginsburg}, {Price-Whelan}, {Kerzendorf}, {Conley}, {Crighton}, {Barbary}, {Muna}, {Ferguson}, {Grollier}, {Parikh}, {Nair}, {Unther}, {Deil}, {Woillez}, {Conseil}, {Kramer}, {Turner}, {Singer}, {Fox}, {Weaver}, {Zabalza}, {Edwards}, {Azalee Bostroem}, {Burke}, {Casey}, {Crawford}, {Dencheva}, {Ely}, {Jenness}, {Labrie}, {Lim}, {Pierfederici}, {Pontzen}, {Ptak}, {Refsdal}, {Servillat}, \& {Streicher}}]{Astropy2013}
{Astropy Collaboration}, {Robitaille}, T.~P., {Tollerud}, E.~J., {et~al.} 2013, \aap, 558, A33, \dodoi{10.1051/0004-6361/201322068}

\bibitem[{{Astropy Collaboration} {et~al.}(2018){Astropy Collaboration}, {Price-Whelan}, {Sip{\H{o}}cz}, {G{\"u}nther}, {Lim}, {Crawford}, {Conseil}, {Shupe}, {Craig}, {Dencheva}, {Ginsburg}, {Vand erPlas}, {Bradley}, {P{\'e}rez-Su{\'a}rez}, {de Val-Borro}, {Aldcroft}, {Cruz}, {Robitaille}, {Tollerud}, {Ardelean}, {Babej}, {Bach}, {Bachetti}, {Bakanov}, {Bamford}, {Barentsen}, {Barmby}, {Baumbach}, {Berry}, {Biscani}, {Boquien}, {Bostroem}, {Bouma}, {Brammer}, {Bray}, {Breytenbach}, {Buddelmeijer}, {Burke}, {Calderone}, {Cano Rodr{\'\i}guez}, {Cara}, {Cardoso}, {Cheedella}, {Copin}, {Corrales}, {Crichton}, {D'Avella}, {Deil}, {Depagne}, {Dietrich}, {Donath}, {Droettboom}, {Earl}, {Erben}, {Fabbro}, {Ferreira}, {Finethy}, {Fox}, {Garrison}, {Gibbons}, {Goldstein}, {Gommers}, {Greco}, {Greenfield}, {Groener}, {Grollier}, {Hagen}, {Hirst}, {Homeier}, {Horton}, {Hosseinzadeh}, {Hu}, {Hunkeler}, {Ivezi{\'c}}, {Jain}, {Jenness}, {Kanarek}, {Kendrew}, {Kern}, {Kerzendorf}, {Khvalko}, {King}, {Kirkby}, {Kulkarni},
  {Kumar}, {Lee}, {Lenz}, {Littlefair}, {Ma}, {Macleod}, {Mastropietro}, {McCully}, {Montagnac}, {Morris}, {Mueller}, {Mumford}, {Muna}, {Murphy}, {Nelson}, {Nguyen}, {Ninan}, {N{\"o}the}, {Ogaz}, {Oh}, {Parejko}, {Parley}, {Pascual}, {Patil}, {Patil}, {Plunkett}, {Prochaska}, {Rastogi}, {Reddy Janga}, {Sabater}, {Sakurikar}, {Seifert}, {Sherbert}, {Sherwood-Taylor}, {Shih}, {Sick}, {Silbiger}, {Singanamalla}, {Singer}, {Sladen}, {Sooley}, {Sornarajah}, {Streicher}, {Teuben}, {Thomas}, {Tremblay}, {Turner}, {Terr{\'o}n}, {van Kerkwijk}, {de la Vega}, {Watkins}, {Weaver}, {Whitmore}, {Woillez}, {Zabalza}, \& {Astropy Contributors}}]{Astropy2018}
{Astropy Collaboration}, {Price-Whelan}, A.~M., {Sip{\H{o}}cz}, B.~M., {et~al.} 2018, \aj, 156, 123, \dodoi{10.3847/1538-3881/aabc4f}

\bibitem[{{Bertoldi} {et~al.}(1999){Bertoldi}, {Timmermann}, {Rosenthal}, {Drapatz}, \& {Wright}}]{Bertoldi1999}
{Bertoldi}, F., {Timmermann}, R., {Rosenthal}, D., {Drapatz}, S., \& {Wright}, C.~M. 1999, \aap, 346, 267, \dodoi{10.48550/arXiv.astro-ph/9904261}

\bibitem[{{Bokeh Development Team}(2018)}]{Bokeh2018}
{Bokeh Development Team}. 2018, Bokeh: Python library for interactive visualization.
\newblock \url{https://bokeh.pydata.org/en/latest/}

\bibitem[{{Brandl} {et~al.}(2006){Brandl}, {Bernard-Salas}, {Spoon}, {Devost}, {Sloan}, {Guilles}, {Wu}, {Houck}, {Weedman}, {Armus}, {Appleton}, {Soifer}, {Charmandaris}, {Hao}, {Higdon}, {Marshall}, \& {Herter}}]{Brandl2006}
{Brandl}, B.~R., {Bernard-Salas}, J., {Spoon}, H.~W.~W., {et~al.} 2006, \apj, 653, 1129, \dodoi{10.1086/508849}

\bibitem[{{Calzetti} {et~al.}(1994){Calzetti}, {Kinney}, \& {Storchi-Bergmann}}]{Calzetti1994}
{Calzetti}, D., {Kinney}, A.~L., \& {Storchi-Bergmann}, T. 1994, \apj, 429, 582, \dodoi{10.1086/174346}

\bibitem[{{Cardelli} {et~al.}(1989){Cardelli}, {Clayton}, \& {Mathis}}]{Cardelli1989}
{Cardelli}, J.~A., {Clayton}, G.~C., \& {Mathis}, J.~S. 1989, \apj, 345, 245, \dodoi{10.1086/167900}

\bibitem[{{Chapman} {et~al.}(2009){Chapman}, {Mundy}, {Lai}, \& {Evans}}]{Chapman2009}
{Chapman}, N.~L., {Mundy}, L.~G., {Lai}, S.-P., \& {Evans}, Neal~J., I. 2009, \apj, 690, 496, \dodoi{10.1088/0004-637X/690/1/496}

\bibitem[{{Chiar} \& {Tielens}(2006)}]{Chiar2006}
{Chiar}, J.~E., \& {Tielens}, A.~G.~G.~M. 2006, \apj, 637, 774, \dodoi{10.1086/498406}

\bibitem[{{Corrales} {et~al.}(2016){Corrales}, {Garc{\'\i}a}, {Wilms}, \& {Baganoff}}]{Corrales2016}
{Corrales}, L.~R., {Garc{\'\i}a}, J., {Wilms}, J., \& {Baganoff}, F. 2016, \mnras, 458, 1345, \dodoi{10.1093/mnras/stw376}

\bibitem[{{Decleir} {et~al.}(2022){Decleir}, {Gordon}, {Andrews}, {Clayton}, {Cushing}, {Misselt}, {Pendleton}, {Rayner}, {Vacca}, \& {Whittet}}]{Decleir2022}
{Decleir}, M., {Gordon}, K.~D., {Andrews}, J.~E., {et~al.} 2022, \apj, 930, 15, \dodoi{10.3847/1538-4357/ac5dbe}

\bibitem[{{Desert} {et~al.}(1990){Desert}, {Boulanger}, \& {Puget}}]{Desert1990}
{Desert}, F.~X., {Boulanger}, F., \& {Puget}, J.~L. 1990, \aap, 237, 215

\bibitem[{{Disney} {et~al.}(1989){Disney}, {Davies}, \& {Phillipps}}]{Disney1989}
{Disney}, M., {Davies}, J., \& {Phillipps}, S. 1989, \mnras, 239, 939, \dodoi{10.1093/mnras/239.3.939}

\bibitem[{{Dom{\'\i}nguez} {et~al.}(2013){Dom{\'\i}nguez}, {Siana}, {Henry}, {Scarlata}, {Bedregal}, {Malkan}, {Atek}, {Ross}, {Colbert}, {Teplitz}, {Rafelski}, {McCarthy}, {Bunker}, {Hathi}, {Dressler}, {Martin}, \& {Masters}}]{Dominguez2013}
{Dom{\'\i}nguez}, A., {Siana}, B., {Henry}, A.~L., {et~al.} 2013, \apj, 763, 145, \dodoi{10.1088/0004-637X/763/2/145}

\bibitem[{{Draine}(2003)}]{Draine2003}
{Draine}, B.~T. 2003, \araa, 41, 241, \dodoi{10.1146/annurev.astro.41.011802.094840}

\bibitem[{{Draine} \& {Li}(2007)}]{Draine2007}
{Draine}, B.~T., \& {Li}, A. 2007, \apj, 657, 810, \dodoi{10.1086/511055}

\bibitem[{{Fitzpatrick} \& {Massa}(2009)}]{Fitzpatrick2009}
{Fitzpatrick}, E.~L., \& {Massa}, D. 2009, \apj, 699, 1209, \dodoi{10.1088/0004-637X/699/2/1209}

\bibitem[{{Fitzpatrick} {et~al.}(2019){Fitzpatrick}, {Massa}, {Gordon}, {Bohlin}, \& {Clayton}}]{Fitzpatrick2019}
{Fitzpatrick}, E.~L., {Massa}, D., {Gordon}, K.~D., {Bohlin}, R., \& {Clayton}, G.~C. 2019, \apj, 886, 108, \dodoi{10.3847/1538-4357/ab4c3a}

\bibitem[{{Flaherty} {et~al.}(2007){Flaherty}, {Pipher}, {Megeath}, {Winston}, {Gutermuth}, {Muzerolle}, {Allen}, \& {Fazio}}]{Flaherty2007}
{Flaherty}, K.~M., {Pipher}, J.~L., {Megeath}, S.~T., {et~al.} 2007, \apj, 663, 1069, \dodoi{10.1086/518411}

\bibitem[{{Foreman-Mackey} {et~al.}(2013){Foreman-Mackey}, {Hogg}, {Lang}, \& {Goodman}}]{Foreman-Mackey2013}
{Foreman-Mackey}, D., {Hogg}, D.~W., {Lang}, D., \& {Goodman}, J. 2013, \pasp, 125, 306, \dodoi{10.1086/670067}

\bibitem[{{Fritz} {et~al.}(2011){Fritz}, {Gillessen}, {Dodds-Eden}, {Lutz}, {Genzel}, {Raab}, {Ott}, {Pfuhl}, {Eisenhauer}, \& {Yusef-Zadeh}}]{Fritz2011}
{Fritz}, T.~K., {Gillessen}, S., {Dodds-Eden}, K., {et~al.} 2011, \apj, 737, 73, \dodoi{10.1088/0004-637X/737/2/73}

\bibitem[{{Gao} {et~al.}(2009){Gao}, {Jiang}, \& {Li}}]{Gao2009}
{Gao}, J., {Jiang}, B.~W., \& {Li}, A. 2009, \apj, 707, 89, \dodoi{10.1088/0004-637X/707/1/89}

\bibitem[{{Genzel} {et~al.}(1998){Genzel}, {Lutz}, {Sturm}, {Egami}, {Kunze}, {Moorwood}, {Rigopoulou}, {Spoon}, {Sternberg}, {Tacconi-Garman}, {Tacconi}, \& {Thatte}}]{Genzel1998}
{Genzel}, R., {Lutz}, D., {Sturm}, E., {et~al.} 1998, \apj, 498, 579, \dodoi{10.1086/305576}

\bibitem[{{Gordon} {et~al.}(2023){Gordon}, {Clayton}, {Decleir}, {Fitzpatrick}, {Massa}, {Misselt}, \& {Tollerud}}]{Gordon2023}
{Gordon}, K.~D., {Clayton}, G.~C., {Decleir}, M., {et~al.} 2023, \apj, 950, 86, \dodoi{10.3847/1538-4357/accb59}

\bibitem[{{Gordon} {et~al.}(2021){Gordon}, {Misselt}, {Bouwman}, {Clayton}, {Decleir}, {Hines}, {Pendleton}, {Rieke}, {Smith}, \& {Whittet}}]{Gordon2021}
{Gordon}, K.~D., {Misselt}, K.~A., {Bouwman}, J., {et~al.} 2021, \apj, 916, 33, \dodoi{10.3847/1538-4357/ac00b7}

\bibitem[{{Hensley} \& {Draine}(2020)}]{Hensley2020}
{Hensley}, B.~S., \& {Draine}, B.~T. 2020, \apj, 895, 38, \dodoi{10.3847/1538-4357/ab8cc3}

\bibitem[{{Hensley} \& {Draine}(2021)}]{Hensley2021}
---. 2021, \apj, 906, 73, \dodoi{10.3847/1538-4357/abc8f1}

\bibitem[{{Hern{\'a}n-Caballero} {et~al.}(2016){Hern{\'a}n-Caballero}, {Spoon}, {Lebouteiller}, {Rupke}, \& {Barry}}]{Hernan-Caballero2016}
{Hern{\'a}n-Caballero}, A., {Spoon}, H.~W.~W., {Lebouteiller}, V., {Rupke}, D.~S.~N., \& {Barry}, D.~P. 2016, \mnras, 455, 1796, \dodoi{10.1093/mnras/stv2464}

\bibitem[{{Hern{\'a}n-Caballero} {et~al.}(2020){Hern{\'a}n-Caballero}, {Spoon}, {Alonso-Herrero}, {Hatziminaoglou}, {Magdis}, {P{\'e}rez-Gonz{\'a}lez}, {Pereira-Santaella}, {Arribas}, {Cortzen}, {Labiano}, {Piqueras}, \& {Rigopoulou}}]{Hernan-Caballero2020}
{Hern{\'a}n-Caballero}, A., {Spoon}, H. W.~W., {Alonso-Herrero}, A., {et~al.} 2020, \mnras, 497, 4614, \dodoi{10.1093/mnras/staa2282}

\bibitem[{{Hummer} \& {Storey}(1987)}]{Hummer1987}
{Hummer}, D.~G., \& {Storey}, P.~J. 1987, \mnras, 224, 801, \dodoi{10.1093/mnras/224.3.801}

\bibitem[{Hunter(2007)}]{Hunter2007}
Hunter, J.~D. 2007, Computing in Science \& Engineering, 9, 90, \dodoi{10.1109/MCSE.2007.55}

\bibitem[{{Indebetouw} {et~al.}(2005){Indebetouw}, {Mathis}, {Babler}, {Meade}, {Watson}, {Whitney}, {Wolff}, {Wolfire}, {Cohen}, {Bania}, {Benjamin}, {Clemens}, {Dickey}, {Jackson}, {Kobulnicky}, {Marston}, {Mercer}, {Stauffer}, {Stolovy}, \& {Churchwell}}]{Indebetouw2005}
{Indebetouw}, R., {Mathis}, J.~S., {Babler}, B.~L., {et~al.} 2005, \apj, 619, 931, \dodoi{10.1086/426679}

\bibitem[{{Jiang} {et~al.}(2006){Jiang}, {Gao}, {Omont}, {Schuller}, \& {Simon}}]{Jiang2006}
{Jiang}, B.~W., {Gao}, J., {Omont}, A., {Schuller}, F., \& {Simon}, G. 2006, \aap, 446, 551, \dodoi{10.1051/0004-6361:20053501}

\bibitem[{{Jones} {et~al.}(2013){Jones}, {Fanciullo}, {K{\"o}hler}, {Verstraete}, {Guillet}, {Bocchio}, \& {Ysard}}]{Jones2013}
{Jones}, A.~P., {Fanciullo}, L., {K{\"o}hler}, M., {et~al.} 2013, \aap, 558, A62, \dodoi{10.1051/0004-6361/201321686}

\bibitem[{{Kemper} {et~al.}(2004){Kemper}, {Vriend}, \& {Tielens}}]{Kemper2004}
{Kemper}, F., {Vriend}, W.~J., \& {Tielens}, A.~G.~G.~M. 2004, \apj, 609, 826, \dodoi{10.1086/421339}

\bibitem[{{Kennicutt}(1992)}]{Kennicutt1992}
{Kennicutt}, Robert~C., J. 1992, \apj, 388, 310, \dodoi{10.1086/171154}

\bibitem[{{Kriek} \& {Conroy}(2013)}]{Kriek2013}
{Kriek}, M., \& {Conroy}, C. 2013, \apjl, 775, L16, \dodoi{10.1088/2041-8205/775/1/L16}

\bibitem[{{Lai} {et~al.}(2020){Lai}, {Smith}, {Baba}, {Spoon}, \& {Imanishi}}]{Lai2020}
{Lai}, T. S.~Y., {Smith}, J.~D.~T., {Baba}, S., {Spoon}, H. W.~W., \& {Imanishi}, M. 2020, \apj, 905, 55, \dodoi{10.3847/1538-4357/abc002}

\bibitem[{{Laurent} {et~al.}(2000){Laurent}, {Mirabel}, {Charmandaris}, {Gallais}, {Madden}, {Sauvage}, {Vigroux}, \& {Cesarsky}}]{Laurent2000}
{Laurent}, O., {Mirabel}, I.~F., {Charmandaris}, V., {et~al.} 2000, \aap, 359, 887, \dodoi{10.48550/arXiv.astro-ph/0005376}

\bibitem[{{Li} \& {Chen}(2023)}]{Li2023}
{Li}, J., \& {Chen}, X. 2023, Universe, 9, 364, \dodoi{10.3390/universe9080364}

\bibitem[{{Lutz}(1999)}]{Lutz1999}
{Lutz}, D. 1999, in ESA Special Publication, Vol. 427, The Universe as Seen by ISO, ed. P.~{Cox} \& M.~{Kessler}, 623

\bibitem[{{Lutz} {et~al.}(1996){Lutz}, {Feuchtgruber}, {Genzel}, {Kunze}, {Rigopoulou}, {Spoon}, {Wright}, {Egami}, {Katterloher}, {Sturm}, {Wieprecht}, {Sternberg}, {Moorwood}, \& {de Graauw}}]{Lutz1996}
{Lutz}, D., {Feuchtgruber}, H., {Genzel}, R., {et~al.} 1996, \aap, 315, L269

\bibitem[{{Markov} {et~al.}(2023){Markov}, {Gallerani}, {Pallottini}, {Sommovigo}, {Carniani}, {Ferrara}, {Parlanti}, \& {Di Mascia}}]{Markov2023}
{Markov}, V., {Gallerani}, S., {Pallottini}, A., {et~al.} 2023, arXiv e-prints, arXiv:2304.11178, \dodoi{10.48550/arXiv.2304.11178}

\bibitem[{{Massa} {et~al.}(1983){Massa}, {Savage}, \& {Fitzpatrick}}]{Massa1983}
{Massa}, D., {Savage}, B.~D., \& {Fitzpatrick}, E.~L. 1983, \apj, 266, 662, \dodoi{10.1086/160813}

\bibitem[{{Mathis}(1994)}]{Mathis1994}
{Mathis}, J.~S. 1994, \apj, 422, 176, \dodoi{10.1086/173715}

\bibitem[{{Mathis} {et~al.}(1977){Mathis}, {Rumpl}, \& {Nordsieck}}]{Mathis1977}
{Mathis}, J.~S., {Rumpl}, W., \& {Nordsieck}, K.~H. 1977, \apj, 217, 425, \dodoi{10.1086/155591}

\bibitem[{{McClure}(2009)}]{McClure2009}
{McClure}, M. 2009, \apjl, 693, L81, \dodoi{10.1088/0004-637X/693/2/L81}

\bibitem[{{Moustakas} {et~al.}(2006){Moustakas}, {Kennicutt}, \& {Tremonti}}]{Moustakas2006}
{Moustakas}, J., {Kennicutt}, Robert~C., J., \& {Tremonti}, C.~A. 2006, \apj, 642, 775, \dodoi{10.1086/500964}

\bibitem[{{Narayanan} {et~al.}(2018){Narayanan}, {Conroy}, {Dav{\'e}}, {Johnson}, \& {Popping}}]{Narayanan2018}
{Narayanan}, D., {Conroy}, C., {Dav{\'e}}, R., {Johnson}, B.~D., \& {Popping}, G. 2018, \apj, 869, 70, \dodoi{10.3847/1538-4357/aaed25}

\bibitem[{{Nishiyama} {et~al.}(2009){Nishiyama}, {Tamura}, {Hatano}, {Kato}, {Tanab{\'e}}, {Sugitani}, \& {Nagata}}]{Nishiyama2009}
{Nishiyama}, S., {Tamura}, M., {Hatano}, H., {et~al.} 2009, \apj, 696, 1407, \dodoi{10.1088/0004-637X/696/2/1407}

\bibitem[{{Nishiyama} {et~al.}(2006){Nishiyama}, {Nagata}, {Kusakabe}, {Matsunaga}, {Naoi}, {Kato}, {Nagashima}, {Sugitani}, {Tamura}, {Tanab{\'e}}, \& {Sato}}]{Nishiyama2006}
{Nishiyama}, S., {Nagata}, T., {Kusakabe}, N., {et~al.} 2006, \apj, 638, 839, \dodoi{10.1086/499038}

\bibitem[{{Ossenkopf} {et~al.}(1992){Ossenkopf}, {Henning}, \& {Mathis}}]{Ossenkopf1992}
{Ossenkopf}, V., {Henning}, T., \& {Mathis}, J.~S. 1992, \aap, 261, 567

\bibitem[{{Reddy} {et~al.}(2015){Reddy}, {Kriek}, {Shapley}, {Freeman}, {Siana}, {Coil}, {Mobasher}, {Price}, {Sanders}, \& {Shivaei}}]{Reddy2015}
{Reddy}, N.~A., {Kriek}, M., {Shapley}, A.~E., {et~al.} 2015, \apj, 806, 259, \dodoi{10.1088/0004-637X/806/2/259}

\bibitem[{{Rieke} \& {Lebofsky}(1985)}]{Rieke1985}
{Rieke}, G.~H., \& {Lebofsky}, M.~J. 1985, \apj, 288, 618, \dodoi{10.1086/162827}

\bibitem[{{Rigopoulou} {et~al.}(1999){Rigopoulou}, {Spoon}, {Genzel}, {Lutz}, {Moorwood}, \& {Tran}}]{Rigopoulou1999}
{Rigopoulou}, D., {Spoon}, H.~W.~W., {Genzel}, R., {et~al.} 1999, \aj, 118, 2625, \dodoi{10.1086/301146}

\bibitem[{{Rosenthal} {et~al.}(2000){Rosenthal}, {Bertoldi}, \& {Drapatz}}]{Rosenthal2000}
{Rosenthal}, D., {Bertoldi}, F., \& {Drapatz}, S. 2000, \aap, 356, 705, \dodoi{10.48550/arXiv.astro-ph/0002456}

\bibitem[{{Salim} \& {Narayanan}(2020)}]{Salim2020}
{Salim}, S., \& {Narayanan}, D. 2020, \araa, 58, 529, \dodoi{10.1146/annurev-astro-032620-021933}

\bibitem[{{Salim} {et~al.}(2016){Salim}, {Lee}, {Janowiecki}, {da Cunha}, {Dickinson}, {Boquien}, {Burgarella}, {Salzer}, \& {Charlot}}]{Salim2016}
{Salim}, S., {Lee}, J.~C., {Janowiecki}, S., {et~al.} 2016, \apjs, 227, 2, \dodoi{10.3847/0067-0049/227/1/2}

\bibitem[{{Shivaei} {et~al.}(2015){Shivaei}, {Reddy}, {Shapley}, {Kriek}, {Siana}, {Mobasher}, {Coil}, {Freeman}, {Sanders}, {Price}, {de Groot}, \& {Azadi}}]{Shivaei2015}
{Shivaei}, I., {Reddy}, N.~A., {Shapley}, A.~E., {et~al.} 2015, \apj, 815, 98, \dodoi{10.1088/0004-637X/815/2/98}

\bibitem[{{Sirocky} {et~al.}(2008){Sirocky}, {Levenson}, {Elitzur}, {Spoon}, \& {Armus}}]{Sirocky2008}
{Sirocky}, M.~M., {Levenson}, N.~A., {Elitzur}, M., {Spoon}, H.~W.~W., \& {Armus}, L. 2008, \apj, 678, 729, \dodoi{10.1086/586727}

\bibitem[{{Smith} {et~al.}(2007){Smith}, {Draine}, {Dale}, {Moustakas}, {Kennicutt}, {Helou}, {Armus}, {Roussel}, {Sheth}, {Bendo}, {Buckalew}, {Calzetti}, {Engelbracht}, {Gordon}, {Hollenbach}, {Li}, {Malhotra}, {Murphy}, \& {Walter}}]{Smith2007}
{Smith}, J.~D.~T., {Draine}, B.~T., {Dale}, D.~A., {et~al.} 2007, \apj, 656, 770, \dodoi{10.1086/510549}

\bibitem[{{Spoon} {et~al.}(2004){Spoon}, {Armus}, {Cami}, {Tielens}, {Chiar}, {Peeters}, {Keane}, {Charmandaris}, {Appleton}, {Teplitz}, \& {Burgdorf}}]{Spoon2004}
{Spoon}, H.~W.~W., {Armus}, L., {Cami}, J., {et~al.} 2004, \apjs, 154, 184, \dodoi{10.1086/422813}

\bibitem[{{Spoon} {et~al.}(2006){Spoon}, {Tielens}, {Armus}, {Sloan}, {Sargent}, {Cami}, {Charmandaris}, {Houck}, \& {Soifer}}]{Spoon2006}
{Spoon}, H.~W.~W., {Tielens}, A.~G.~G.~M., {Armus}, L., {et~al.} 2006, \apj, 638, 759, \dodoi{10.1086/498566}

\bibitem[{{Spoon} {et~al.}(2022){Spoon}, {Hern{\'a}n-Caballero}, {Rupke}, {Waters}, {Lebouteiller}, {Tielens}, {Loredo}, {Su}, \& {Viola}}]{Spoon2022}
{Spoon}, H.~W.~W., {Hern{\'a}n-Caballero}, A., {Rupke}, D., {et~al.} 2022, \apjs, 259, 37, \dodoi{10.3847/1538-4365/ac4989}

\bibitem[{{Stecher}(1965)}]{Stecher1965}
{Stecher}, T.~P. 1965, \apj, 142, 1683, \dodoi{10.1086/148462}

\bibitem[{{Steglich} {et~al.}(2010){Steglich}, {J{\"a}ger}, {Rouill{\'e}}, {Huisken}, {Mutschke}, \& {Henning}}]{Steglich2010}
{Steglich}, M., {J{\"a}ger}, C., {Rouill{\'e}}, G., {et~al.} 2010, \apjl, 712, L16, \dodoi{10.1088/2041-8205/712/1/L16}

\bibitem[{{Storey} \& {Hummer}(1995)}]{Storey1995}
{Storey}, P.~J., \& {Hummer}, D.~G. 1995, \mnras, 272, 41, \dodoi{10.1093/mnras/272.1.41}

\bibitem[{{Tran} {et~al.}(2001){Tran}, {Lutz}, {Genzel}, {Rigopoulou}, {Spoon}, {Sturm}, {Gerin}, {Hines}, {Moorwood}, {Sanders}, {Scoville}, {Taniguchi}, \& {Ward}}]{Tran2001}
{Tran}, Q.~D., {Lutz}, D., {Genzel}, R., {et~al.} 2001, \apj, 552, 527, \dodoi{10.1086/320543}

\bibitem[{{van Breemen} {et~al.}(2011){van Breemen}, {Min}, {Chiar}, {Waters}, {Kemper}, {Boogert}, {Cami}, {Decin}, {Knez}, {Sloan}, \& {Tielens}}]{vanBreemen2011}
{van Breemen}, J.~M., {Min}, M., {Chiar}, J.~E., {et~al.} 2011, \aap, 526, A152, \dodoi{10.1051/0004-6361/200811142}

\bibitem[{{van der Walt} {et~al.}(2011){van der Walt}, {Colbert}, \& {Varoquaux}}]{VanderWalt2011}
{van der Walt}, S., {Colbert}, S.~C., \& {Varoquaux}, G. 2011, Computing in Science Engineering, 13, 22

\bibitem[{{Virtanen} {et~al.}(2020){Virtanen}, {Gommers}, {Oliphant}, {Haberland}, {Reddy}, {Cournapeau}, {Burovski}, {Peterson}, {Weckesser}, {Bright}, {van der Walt}, {Brett}, {Wilson}, {Jarrod Millman}, {Mayorov}, {Nelson}, {Jones}, {Kern}, {Larson}, {Carey}, {Polat}, {Feng}, {Moore}, {Vand erPlas}, {Laxalde}, {Perktold}, {Cimrman}, {Henriksen}, {Quintero}, {Harris}, {Archibald}, {Ribeiro}, {Pedregosa}, {van Mulbregt}, \& {Contributors}}]{Virtanen2020}
{Virtanen}, P., {Gommers}, R., {Oliphant}, T.~E., {et~al.} 2020, Nature Methods, 17, 261, \dodoi{https://doi.org/10.1038/s41592-019-0686-2}

\bibitem[{{Wang} {et~al.}(2013){Wang}, {Gao}, {Jiang}, {Li}, \& {Chen}}]{Wang2013}
{Wang}, S., {Gao}, J., {Jiang}, B.~W., {Li}, A., \& {Chen}, Y. 2013, \apj, 773, 30, \dodoi{10.1088/0004-637X/773/1/30}

\bibitem[{{Wang} {et~al.}(2015){Wang}, {Li}, \& {Jiang}}]{Wang2015}
{Wang}, S., {Li}, A., \& {Jiang}, B.~W. 2015, \apj, 811, 38, \dodoi{10.1088/0004-637X/811/1/38}

\bibitem[{{Weingartner} \& {Draine}(2001)}]{Weingartner2001}
{Weingartner}, J.~C., \& {Draine}, B.~T. 2001, \apjs, 134, 263, \dodoi{10.1086/320852}

\bibitem[{{Witstok} {et~al.}(2023){Witstok}, {Shivaei}, {Smit}, {Maiolino}, {Carniani}, {Curtis-Lake}, {Ferruit}, {Arribas}, {Bunker}, {Cameron}, {Charlot}, {Chevallard}, {Curti}, {de Graaff}, {D'Eugenio}, {Giardino}, {Looser}, {Rawle}, {Rodr{\'\i}guez del Pino}, {Willott}, {Alberts}, {Baker}, {Boyett}, {Egami}, {Eisenstein}, {Endsley}, {Hainline}, {Ji}, {Johnson}, {Kumari}, {Lyu}, {Nelson}, {Perna}, {Rieke}, {Robertson}, {Sandles}, {Saxena}, {Scholtz}, {Sun}, {Tacchella}, {Williams}, \& {Willmer}}]{Witstok2023}
{Witstok}, J., {Shivaei}, I., {Smit}, R., {et~al.} 2023, \nat, 621, 267, \dodoi{10.1038/s41586-023-06413-w}

\bibitem[{{Witt} \& {Gordon}(1996)}]{Witt1996}
{Witt}, A.~N., \& {Gordon}, K.~D. 1996, \apj, 463, 681, \dodoi{10.1086/177282}

\bibitem[{{Witt} \& {Gordon}(2000)}]{Witt2000}
---. 2000, \apj, 528, 799, \dodoi{10.1086/308197}

\bibitem[{{Woolf} \& {Ney}(1969)}]{Woolf1969}
{Woolf}, N.~J., \& {Ney}, E.~P. 1969, \apjl, 155, L181, \dodoi{10.1086/180331}

\bibitem[{{Xue} {et~al.}(2016){Xue}, {Jiang}, {Gao}, {Liu}, {Wang}, \& {Li}}]{Xue2016}
{Xue}, M., {Jiang}, B.~W., {Gao}, J., {et~al.} 2016, \apjs, 224, 23, \dodoi{10.3847/0067-0049/224/2/23}

\bibitem[{{Zasowski} {et~al.}(2009){Zasowski}, {Majewski}, {Indebetouw}, {Meade}, {Nidever}, {Patterson}, {Babler}, {Skrutskie}, {Watson}, {Whitney}, \& {Churchwell}}]{Zasowski2009}
{Zasowski}, G., {Majewski}, S.~R., {Indebetouw}, R., {et~al.} 2009, \apj, 707, 510, \dodoi{10.1088/0004-637X/707/1/510}

\end{thebibliography}



\end{CJK*}
\end{document}